\documentclass[
aip,
preprint,
amsmath,
amssymb]{revtex4-1}
\usepackage{graphicx}
\usepackage{hyperref}
\hypersetup{
    colorlinks=true,
    linkcolor=blue,
    filecolor=blue,      
    urlcolor=blue,
    citecolor=blue,
}
\usepackage[usenames,dvipsnames]{color}
\newcommand{\units}[1]{\ensuremath{\,\mathrm{#1}}}

\begin{document}


\title{Supercurrent diode effect in thin film Nb tracks} 



\author{N~Satchell}
\affiliation{School of Physics and Astronomy, University of Leeds, Leeds, LS2 9JT, United Kingdom}

\author{PM~Shepley}
\affiliation{School of Physics and Astronomy, University of Leeds, Leeds, LS2 9JT, United Kingdom}

\author{MC~Rosamond}
\affiliation{School of Electronic and Electrical Engineering, University of Leeds, Leeds LS2 9JT, United Kingdom}

\author{G~Burnell}
\email{g.burnell@leeds.ac.uk}
\affiliation{School of Physics and Astronomy, University of Leeds, Leeds, LS2 9JT, United Kingdom}


\date{\today}

\begin{abstract}
We demonstrate nonreciprocal critical current in 65~nm thick polycrystalline and epitaxial Nb thin films patterned into tracks. The nonreciprocal behavior gives a supercurrent diode effect, where the current passed in one direction is a supercurrent and the other direction is a normal state (resistive) current. We attribute fabrication artefacts to creating the supercurrent diode effect in our tracks. We study the variation of the diode effect with temperature and magnetic field, and find a dependence with the width of the Nb tracks from 2-10~$\mu$m. For both polycrystalline and epitaxial samples, we find that tracks of width 4~$\mu$m provides the largest supercurrent diode efficiency of up to $\approx30\%$, with the effect reducing or disappearing in the widest tracks of 10~$\mu$m. We propose a model based on the limiting contributions to the critical current density to explain the track width dependence of the induced supercurrent diode effect. It is anticipated that the supercurrent diode will become a ubiquitous component of the superconducting computer. 
\end{abstract}


\maketitle 

\section{Introduction}

The supercurrent diode is an analogue to rectification in semiconductor pn junctions, where current is allowed to flow only in one direction. In the supercurrent diode, the critical current of the device ($I_c$) is nonreciprocal ($I_c^+ \neq I_c^-$), leading to the situation where a dissipationless supercurrent can pass in one direction, but upon reversing the current direction, the device becomes resistive. It has been suggested that a supercurrent diode may become a useful component for digital processing in a superconducting computer.

Nonreciprocal $I_c$ behavior has been predicted and observed prior to the common use of the term supercurrent diode effect. Some notable early experimental works include superconducting foils under external magnetic fields\cite{PhysRev.156.412} and reports in superconductor-ferromagnet hybrid systems\cite{doi:10.1063/1.1789231, PhysRevB.72.064509, doi:10.1063/1.2199468, doi:10.1063/1.3009207}.

Theoretically, a nonreciprocal $I_c$ can manifest in Josephson devices with anomalous $\phi_0$ phase differences\cite{PhysRevLett.102.227005, PhysRevB.105.104508} or when the barrier materials have symmetry breaking\cite{PhysRevX.12.041013, davydova2022universal}. A common requirement of such predictions are that the barriers contain exotic materials such as ferromagnetic multilayers\cite{Eschrig_2015, PhysRevLett.98.077003, Heim_2013, PhysRevB.91.214511, PhysRevB.99.224513, PhysRevApplied.14.014003, PhysRevB.106.174502}, topological insulators\cite{PhysRevB.94.134506, PhysRevB.96.165422, Li_2022}, Weyl semimetals\cite{PhysRevB.98.245418, PhysRevB.101.035120, PhysRevB.101.075110, PhysRevB.102.085144}, materials with large spin-orbit coupling\cite{PhysRevB.101.155123, PhysRevB.103.L060503, PhysRevApplied.18.L031001}, or that the electrodes are unconventional superconductors\cite{PhysRevB.63.212502, Liu_2021, xie2022valleypolarized}. Several recent experiments have reported Josephson diode effect in devices with exotic barrier materials\cite{pal2022josephson, baumgartner2022supercurrent, wu2022field, Golod2022Demonstration, jeon2022zero, doi:10.1021/acs.nanolett.2c02899}.

In addition to the Josephson diode effect are reports and predictions for an intrinsic supercurrent diode effect. Theoretically, an intrinsic effect can occur in superconducting material systems with symmetry breaking, such as broken inversion symmetry\cite{daido2022intrinsic, yuan2022supercurrent, PhysRevB.106.224509, PhysRevB.106.L140505, PhysRevB.106.205206} or broken time-reversal symmetry\cite{yuan2022supercurrent, Scammell_2022, wang2022symmetry}. Therefore, much attention has been paid to observations of supercurrent diode effect in exotic materials systems such as multilayers with noncentrosymmetric structure \cite{ando2020observation, wakatsuki2017nonreciprocal, zhang2020nonreciprocal, Narita2022Field}, and systems with time-reversal symmetry breaking\cite{lin2022zero}. In order to fully understand the interesting and novel new physics of such systems, it is important to be able to distinguish between supercurrent diode effect which is intrinsic to the material, and which is induced as a result of device fabrication.

Induced supercurrent diode effect can occur in tracks of $s$-wave superconductors without intrinsic inversion or time-reversal symmetry breaking, such as Nb\cite{hou2022ubiquitous, Suri2022Non, ustavschikov2022diode, gutfreund2023direct}. The induced supercurrent diode effect is attributed to non-perfect device fabrication leading to imperfections in the edges of the tracks. Therefore, the edges of the tracks are unlikely to be identical, which can lead to the unequal generation, penetration and expulsion of vortices on the opposite edges of the tracks, resulting in rectification\cite{PhysRevB.72.172508, hou2022ubiquitous}. The ability to induce a supercurrent diode effect without requirements for exotic materials is useful from a practical point of view, since established industrial processes tend to be based on $s$-wave superconductors such as Nb.

In this work, we report measurements on Nb thin films patterned into tracks of varying widths. From our field and temperature dependent measurements of induced supercurrent diode effect, we report three key experimental observations. Firstly, our Nb tracks are 65$\units{nm}$ thick, refining the thickness in which supercurrent diode effect manifests. Secondly, by varying the width of the tracks, we observe a dependence of the diode effect with track width. Thirdly, we establish that single crystal epitaxy is not a condition for diode effect by measuring both polycrystalline and epitaxial samples. We report our observed diode effect as an efficiency parameter $\eta = \frac{I_c^+ - I_c^-}{I_c^+ + I_c^-}$. Finally, we propose a model to explain our observed track width dependence in terms of the critical current densities in our devices and conclude that induced supercurrent diode effect is a particularly important consideration in track widths where vortex penetration contributes significantly to the overall critical current density.

\section{Methods}

Nb films are deposited by dc magnetron sputtering in the Royce Deposition System \cite{Royce}. The magnetrons are mounted below, and confocal to, the substrate with source-substrate distances of 134~mm. The base pressure of the vacuum chamber is 3$\times10^{-9}\units{mBar}$ with the substrate at room temperature and 1$\times$10$^{-8}\units{mBar}$ with the substrate at 1000$^\circ$C. Nb is grown at a rate of 0.06 nm/s at an Ar (6N purity) gas pressure of 3.6$\times$10$^{-3}\units{mBar}$ to a nominal thickness of 65 nm. The growth rate and film thicknesses are checked by x-ray reflectivity. The first Nb sample was deposited at room temperature on Si/SiO$_x$ substrate and the second Nb sample was deposited at elevated temperature of 1000$^\circ$C onto a single crystal $a$-plane Al$_2$O$_3$ substrate to promote epitaxial growth.

Samples are fabricated into tracks using a direct laser writer to define resist masks in S1813 photoresist, and reactive ion etching at 130 W in a 1:2 Ar:SF6 plasma to etch the Nb films. After fabrication, devices are measured in a continuous flow $^4\mathrm{He}$ cryostat with 3\units{T} horizontal superconducting Helmholtz coils. Traditional 4-point-probe transport geometry is used to measure the current-voltage characteristic of the tracks with combined Keithley 6221-2182A current source and nano-voltmeter in pulse mode with 1~ms pulses and a 1$\%$ duty cycle to avoid a reduced retracking $I_c$ due to heating. Varying the duty cycle from 1-50$\%$ resulted in no changes to the reported $I_c$. The schematic of the fabricated devices and measurement geometry is shown in Figure \ref{Fig1} (a).

The sample grown at room temperature has a superconducting $T_c$ of 8.75 K and a residual-resistivity ratio ($RRR$) of 2.8, giving an estimate for the mean free path ($\ell$) of 6~nm -- indicating a polycrystalline microsctructure. The second sample has a higher $T_c$ of 9.05 K and a $RRR$ of 30, giving an estimate for $\ell$ of 96~nm, consistent with an epitaxial microstructure. The increased $\ell$ is expected for the epitaxial Nb due to the decrease in crystallographic defects such as grain boundaries. The properties of our polycrystalline Nb thin films are reported elsewhere\cite{PhysRevMaterials.4.074801}. 

\section{Results}

\begin{figure}
\includegraphics[width=0.75\columnwidth]{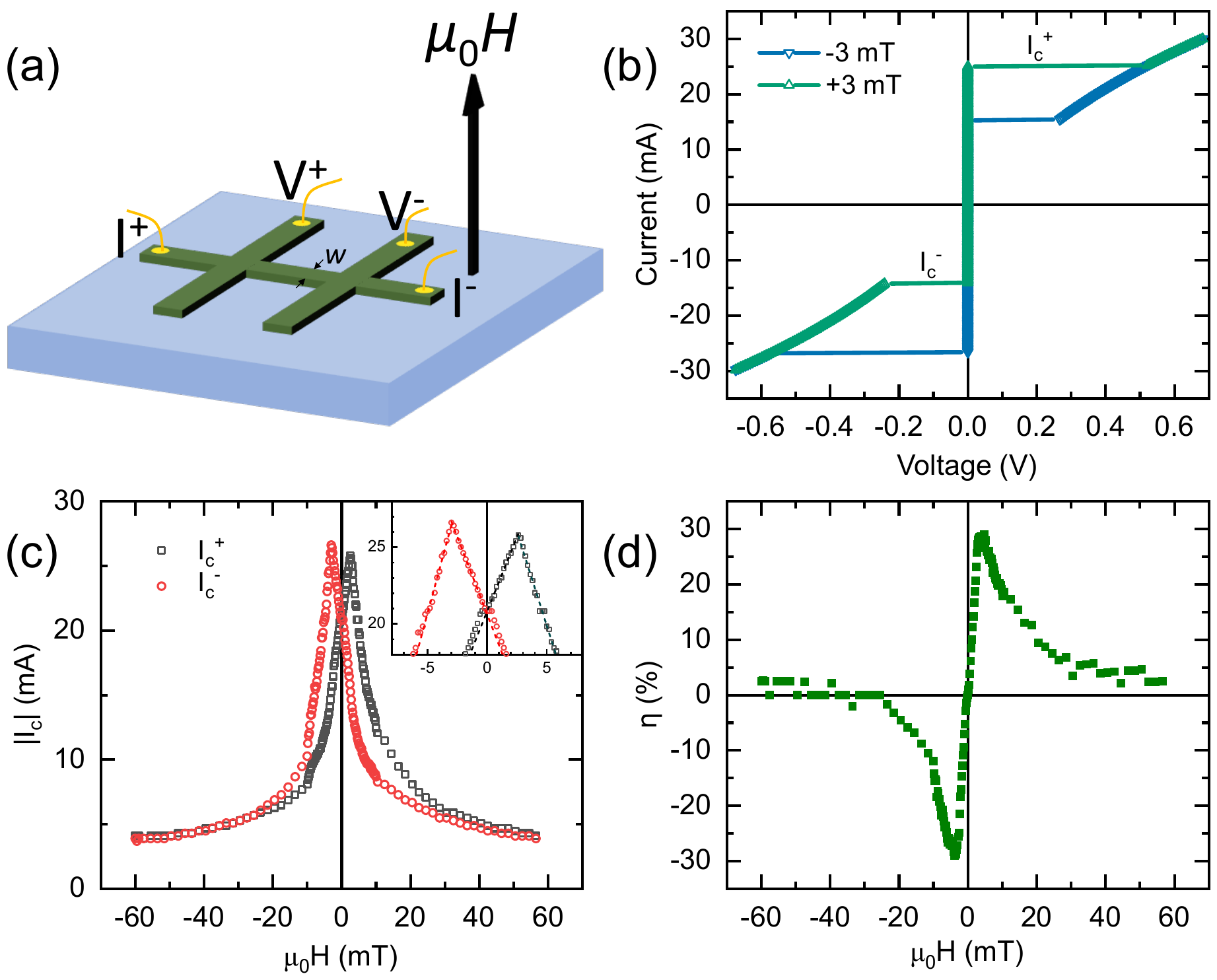}
\caption{\label{Fig1} Supercurrent diode effect in polycrystalline 65 nm thick Nb patterned into a 4 $\mu$m wide track, measured at 5 K. (a) Schematic cross section of the Nb track device showing the applied field and measurement current direction (not to scale). (b) Current-voltage characteristic of the device measured at $\pm$3 mT applied out-of-plane field. (c) Extracted critical currents ($I_c^+$ and $I_c^-$) with out-of-plane applied field, insert shows low field region with fitting to the model described in the text. The uncertainty in determining $I_c$ is the current step size and is smaller than the data points. (d) Diode efficiency $\eta$ corresponding to the dataset in (c). }
\end{figure}

Figure \ref{Fig1} shows the supercurrent diode effect in the polycrystalline sample with 4 $\mu$m wide track. Figure \ref{Fig1} (b) shows the current-voltage ($I-V$) characteristic of our track at applied fields where the diode effect is found to be maximum. $I_c^+$ and $I_c^-$ are extracted from the $I-V$ when the voltage reaches a small threshold value. Large nonreciprocal $I_c^+$ and $I_c^-$ can be seen in the $I-V$ characteristic. When the field polarity is reversed, the nonreciprocal $I_c^+$ and $I_c^-$ are also reversed. In the normal state of the device, the $I-V$ curve shows a slight non-linear dependence, which we attribute to Joule heating as a result of the large current densities. Our measurements at lower current densities (e.g. in wider tracks or at warmer temperatures) show $I-V$ curves following the expected linear metallic behavior.

Figure \ref{Fig1} (c) shows the out-of-plane applied field dependence of $I_c^+$ and $I_c^-$. Due to the diode effect, unexpectedly, the maximum values of $I_c$ do not occur at zero applied field. We attribute this to the vortex penetration field, which is discussed later. The presented field sweep is acquired by sweeping from negative to positive field. A similar curve with the same features is obtained by sweeping from positive to negative field. The sign of the $I_c^+$/$I_c^-$ maximum with positive/negative field does not change with temperature or field sweep direction and appears to favor having the $I_c^+$  ($I_c^-$) maximum in positive (negative) field. Across the 11 samples in our study showing the diode effect, when keeping the device mounting and wiring geometry the same, $I_c^+$ ($I_c^-$) maximum appears in positive (negative) field with a ratio of 8:3. The $I_c^+$/$I_c^-$ maximum with positive/negative field can be reversed by swapping the current wiring direction (Figure \ref{Fig1} (a)). Figure \ref{Fig1} (d) shows the extracted diode efficiency $\eta$, where peak efficiency is achieved around the fields corresponding to $I_c^+$/$I_c^-$ maximum. The maximum in $\eta$ corresponding to the maximum in $I_c^+$ is a common feature of our data, however we believe that this is a consequence of how $\eta$ is defined rather than being a fundamental feature of the induced supercurrent diode. We note a sight asymmetry and digitization in the high field $\eta$ values in this device, Figure \ref{Fig1} (d). The asymmetry in $\eta$ at high fields is present in some of our devices and may indicate that the diode effect persists to high fields in some devices. The digitization of $\eta$ at high fields is a result of determining $I_c$ from the finite current step size in the I--V measurements.

Our interpretation of the origin of the supercurrent diode effect relies upon our samples being in the limit where the critical current is determined by the vorticies in the system, as opposed to being the depairing current. From the $I_c(B)$ dependence, it is possible to extract the maximum super-heating field of the Meissner state\cite{hou2022ubiquitous}, $B_s$. The inset to Figure \ref{Fig1} (c) shows the low field $I_c$ with linear fits (dashed lines). $B_s$ corresponds to the interpolated intercept of the fits to the low field data, when the $B$ offset is taken into account. From the four linear fits shown in Figure \ref{Fig1} (c) inset, we obtain $B_s = 10\pm1$ mT. The expression\cite{hou2022ubiquitous}, $B_s = \phi_0/(\sqrt{3}\pi \xi w$), provides an order of magnitude estimate for $B_s$, where $\phi_0$ is the flux quantum and $\xi$ is the Ginzburg–Landau coherence length ($\xi = 11.6$ nm for polycrystalline thin film Nb\cite{PhysRevMaterials.4.074801}). For $w=4~\mu$m, $B_s =8.2$ mT, in approximate agreement with our experimental findings. This indicates strongly that in the region where the diode effect is observed the critical current is determined by the vorticies.

\begin{figure}
\includegraphics[width=0.75\columnwidth]{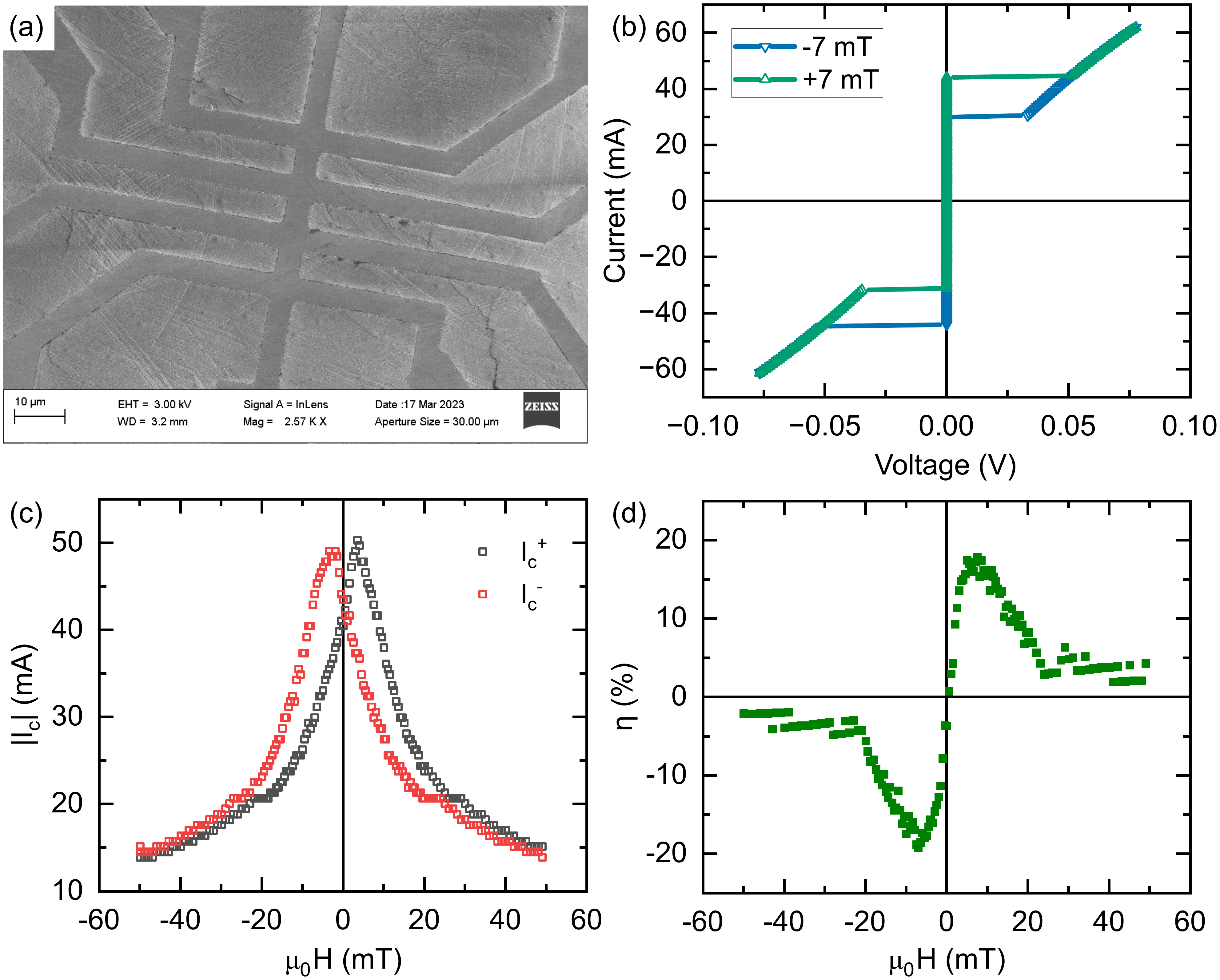}
\caption{\label{Fig6} Supercurrent diode effect in epitaxial 65 nm thick Nb. (a) Scanning electron microscope image of a 4 $\mu$m wide track. (b-d) Electrical transport characteristic of a  4 $\mu$m wide track, measured at 5 K. (a) Current-voltage characteristic of the
device measured at $\pm$7 mT applied out-of-plane field. (b) Extracted critical currents ($I_c^+$ and $I_c^-$) with out-of-plane applied field. The uncertainty in determining $I_c$ is the current step size and is smaller than the data points. (c) Diode efficiency $\eta$ corresponding to the dataset in (b).}
\end{figure}

Figure \ref{Fig6} (a) shows a scanning electron microscope image of the epitaxial 65 nm thick Nb sample patterned into 4 $\mu$m wide track. To avoid charging during imaging, a thin Au layer is deposited over the whole sample. Imaging of multiple tracks reveals the following: firstly, some hardened resist residue remains on the devices at the sidewalls. We do not believe that the resist residue contributes to the observed diode effect; secondly, all our devices have a low line-edge roughness in the range of $10-40$ nm with no significant differences between opposite edges. This implies that the fabrication artefacts believed to be responsible for the induced diode effect are subtle.

Figure \ref{Fig6} shows the supercurrent diode effect in the epitaxial 65 nm thick Nb sample patterned into 4 $\mu$m wide track at 5 K.  From the comparison with the polycrystalline device presented in Figure \ref{Fig1} of the same track width and measurement temperature, the epitaxial device shares many of the same features with the following differences. From the I-V characteristic, Figure \ref{Fig6} (b), the residual resistance of the epitaxial sample is lower than the polycrystalline sample. This sample also displays slightly non-linear dependence, consistent with Joule heating. The tracks patterned from the epitaxial sample has an $I_c$ about twice that of the polycrystalline track, shown in Figure \ref{Fig6} (b) and (c). The increased $I_c$ in the epitaxial sample is most likely due to the combination of increased gap (higher $T_c$), decreased grain boundaries, and increased mean free path compared to the polycrystalline sample. As shown in Figure \ref{Fig6} (d), the measured maximum $\eta$ in this condition is smaller than the polycrystalline sample.

We next study the track width dependence of the supercurrent diode effect in the polycrystalline and epitaxial Nb samples. For polycrystalline Nb, the London penetration depth $\lambda_L = 96$ nm\cite{PhysRevMaterials.4.074801}, providing an estimate of the Pearl penetration depth $\lambda_P = 2\lambda_L^2/t \approx 300$ nm. For comparable epitaxial Nb, $\lambda_L$ tends towards bulk \cite{PhysRevB.72.024506} providing an estimate $\lambda_P \approx 50$ nm. Other works have considered tracks in the limit $w/\lambda_P =1/25$\cite{Suri2022Non} and $w/\lambda_P = 2$\cite{hou2022ubiquitous}, however here we explore a new limit where the width of the tracks are far greater than the Pearl penetration depth, $w>>\lambda_P$, covering between $10 < w/\lambda_P < 200$. In this new limit, we observe a significant superconducting diode effect and report a width dependence in our samples. 

\begin{figure}
\includegraphics[width=0.75\columnwidth]{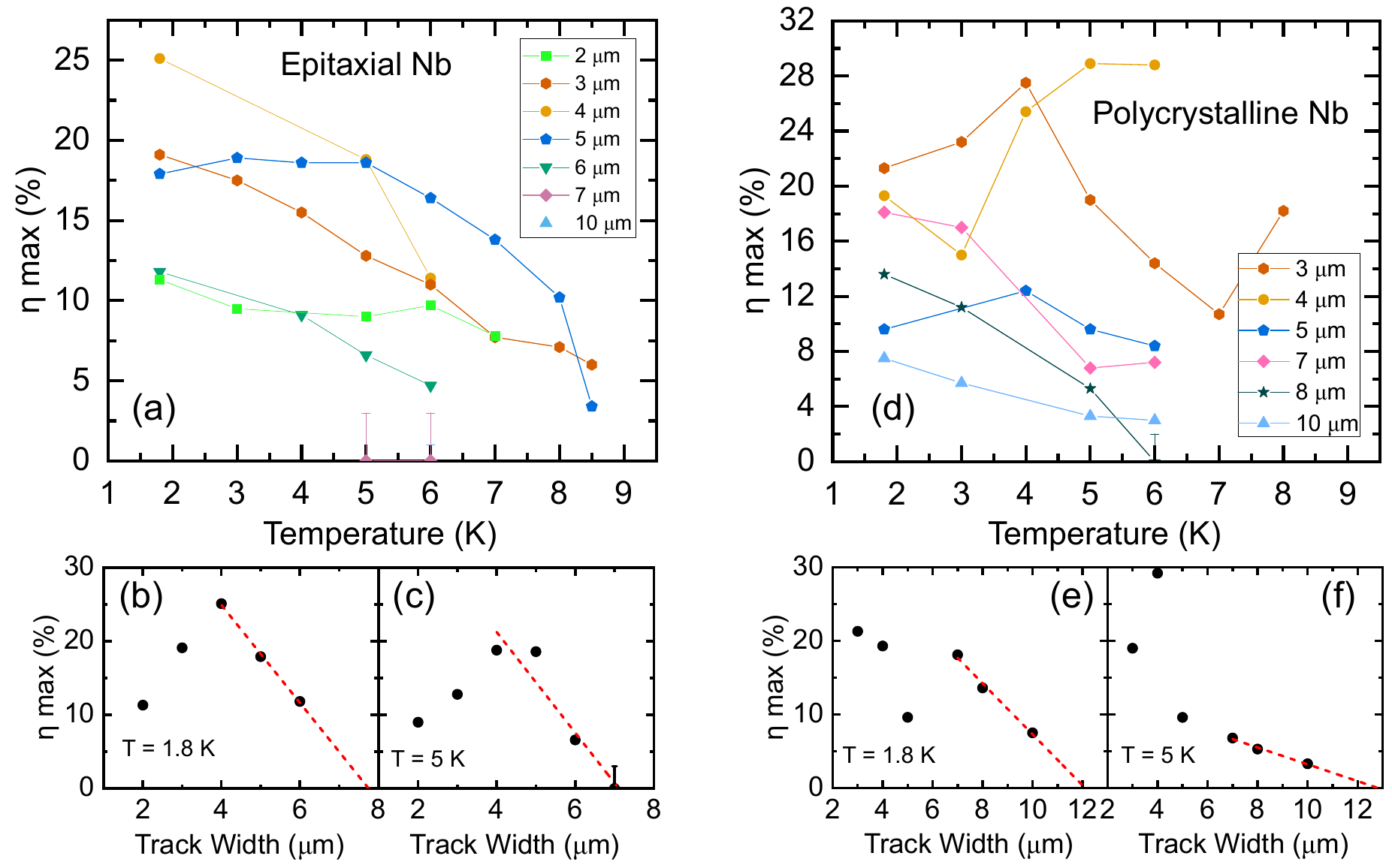}
\caption{\label{Fig2} Supercurrent diode effect in epitaxial and polycrystalline 65 nm thick Nb patterned into tracks (a,d) The maximum diode efficiency parameter, $\eta$, with temperature for a series of tracks with varying width for the epitaxial and polycrystalline samples respectively. (b,c,e,f) Corresponding track width dependence of $\eta$ at (b,e) 1.8 K and (c,f) 5 K. Solid lines in (a,d) represent guides for the eye, while dashed lines in (b,c,e,f) show linear fits to the decay of $\eta$ for the widest tracks.}
\end{figure}

Figure \ref{Fig2} shows the full track width and temperature dependence of the supercurrent diode effect in the epitaxial and polycrystalline Nb samples. Considering first the epitaxial sample, Figure \ref{Fig2} (a), two samples of at $w = 7$ and $10~\mu$m did not show finite $\eta$. In these two tracks, $I_c$ at low temperatures exceeded the maximum output of our current source (100 mA), limiting the range of temperatures we could measure. For the narrower samples which showed $\eta$, the temperature dependence show similar trends for all samples, with the largest $\eta$ at 1.8 K, and $\eta$ decreasing with warming. 

Figure \ref{Fig2} (b) and (c) presents the track width dependence of $\eta$ at fixed temperature. At 1.8 K, a clear peak in $\eta$ is found for the $4~\mu$m track, with $\eta$ decreasing linearly for narrower or wider tracks. At 5 K, the peak in $\eta$ is broader, with $w = 4$ and $5~\mu$m showing similar $\eta$. Again, for narrower or wider tracks $\eta$ decreases linearly with width. Linear fits to the decay of $\eta$ for tracks of $4,5,6$ and $7~\mu$m are presented as dashed lines.

In the polycrystalline Nb samples, Figure \ref{Fig2} (d), tracks in the width regime $3\leq w \leq5~\mu$m follow a similar temperature trend where $\eta$ is largest for temperatures of 4 or 5 K, and decreases from the maximum value as the temperature is cooled or warmed. Tracks with $w \geq7~\mu$m show the largest $\eta$ at the lowest temperature, with $\eta$ decreasing as the temperature is warmed. Considering the track width dependence of $\eta$ at fixed temperature, Figure \ref{Fig2} (e) and (f), $\eta$ shows the largest value for tracks of $w = 3$ or $4~\mu$m. Upon increasing $w$, $\eta$ decreases linearly between $7\leq w \leq10~\mu$m at 1.8 K and between $5\leq w \leq10~\mu$m at 1.8 K. The $5~\mu$m track at 1.8 K is a outlier to this trend. Linear fits to the decay of $\eta$ for tracks of $7,8$ and $10~\mu$m are presented as dashed lines.

\begin{figure}
\includegraphics[width=0.5\columnwidth]{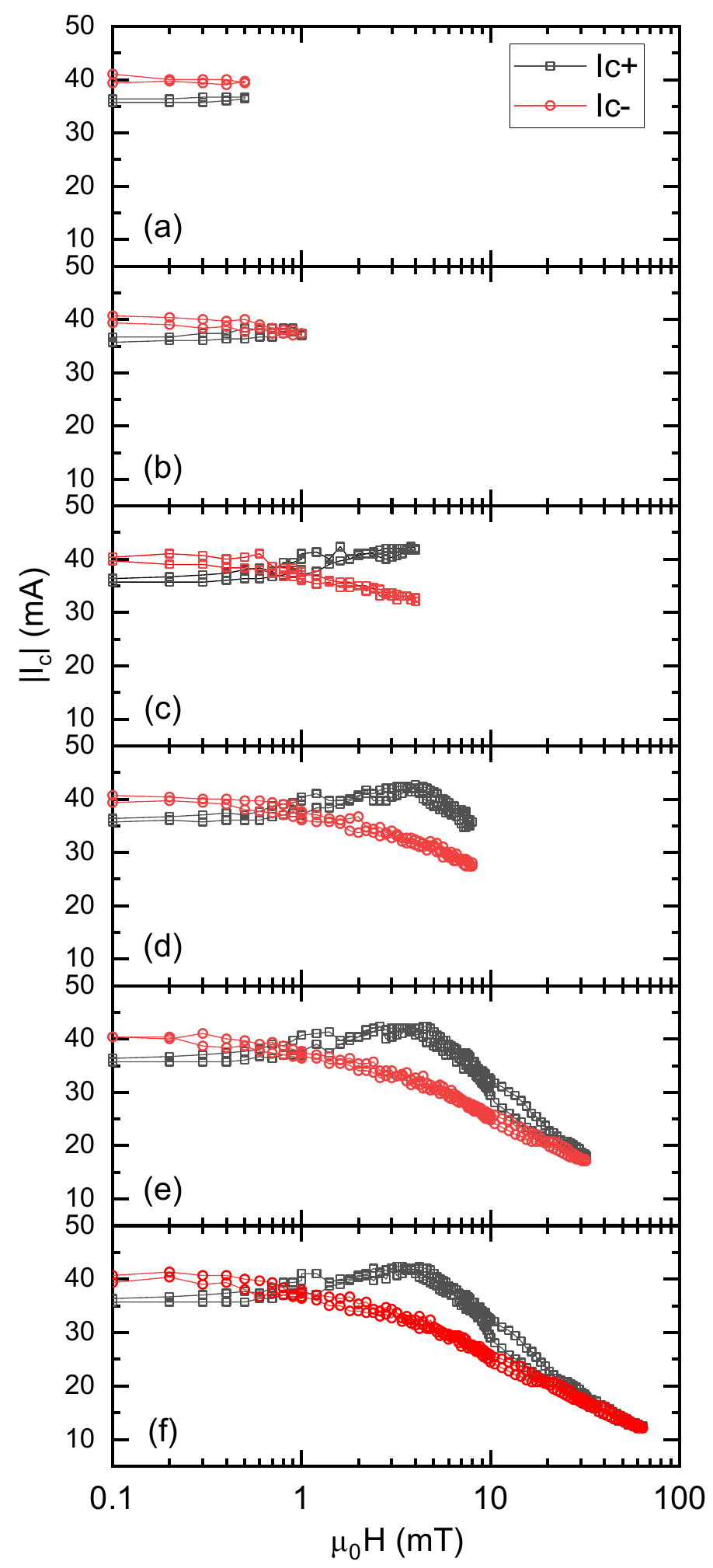}
\caption{\label{Fig4} Initialization of the supercurrent diode effect in epitaxial 65 nm thick Nb patterned into 5 $\mu$m wide track at 6 K. (a-e) $I_c^+$/$I_c^-$ with sequentially larger field sweeps on a semi-log scale. Before each sweep the sample was briefly warmed above $T_c$ and cooled again in zero applied field. The uncertainty in determining $I_c$ is the current step size and is smaller than the data points.}
\end{figure}

In an attempt to further understand the origin of the diode effect and the role of vorticies, we perform initialization experiments on one of our tracks. Figure \ref{Fig4} shows the epitaxial 65 nm thick Nb sample with 5 $\mu$m wide track at 6 K. (a-f) shows the  initial state of the device after zero applied field cool and the conditions necessary to initialize the supercurrent diode effect in the device. At the zero field cooled condition in Figure \ref{Fig4} (a), the device already shows a small diode effect of $\eta=6\%$, presumably due to the small trapped flux in the superconducting magnet. Figure \ref{Fig4} (a-f) present the results of performing sequentially larger field sweeps. Starting from zero applied field, we observe a crossover in $I_c^+$/$I_c^-$ at approximately 1 mT. The crossover at 1 mT, where $\eta=0$, likely corresponds to the true zero field condition. Trapped magnetic flux is a common feature of superconducting coils. Here, we choose not to perform background subtraction of this effect and we present the dataset `as measured'. Due to the larger field step size used to acquire the data presented in Figures \ref{Fig1} and \ref{Fig6}, the small trapped flux is not visible in those datasets. A further consequence of the trapped flux is that our devices will have been cooled through the superconducting transition in a small magnetic field, which will contribute flux vorticies in the as cooled state. The dependence of the diode effect on field cooling was not investigated in this study. The maximum in $I_c^+$ is at about 4 mT. As in Figures \ref{Fig1} and \ref{Fig6}, the maximum diode effect, $\eta$, corresponds to the maximum in $I_c^+$. At larger fields the effect reduces, and $I_c^+$ = $I_c^-$ at about 20 mT. 

In Figure \ref{Fig4}, comparing the out and return field sweep directions, there is a hysteretic behavior in $I_c$, and hence $\eta$. Considering Figure \ref{Fig4} (e), the hysteresis is particularly observable in $I_c^+$ between 10 and 20~mT, but is present over the whole field range where $\eta \neq 0$. The hysteretic behavior with field history suggests that the diode effect is sensitive to the field history of the track. We expect that the preceding field history establishes different vortex states in the track, which therefore influences the magnitude of $\eta$. The largest hysteresis between 10 and 20~mT is reproduced in subsequent larger field sweeps (Figure \ref{Fig4} (f)). 

\section{Discussion}

From the presented experimental data, we can report three key experimental observations. Firstly, at 65 nm thick, our films are in a distinct mid-range thickness with respect to other works in the field \cite{ando2020observation, hou2022ubiquitous}. In this mid-range thickness we have shown that a diode effect is possible. Secondly, by varying the width of the tracks, we observe an unexpected dependence of the diode effect with track width. Thirdly, we extend previous work \cite{hou2022ubiquitous}, which considered only epitaxial Nb and demonstrate that the supercurrent diode effect is present in both polycrystalline and epitaxial Nb.

In our polycrystalline Nb, the coherence length is 11.6 nm \cite{PhysRevMaterials.4.074801}, which is much less than the thickness of the 65 nm film studied here. Our results therefore suggest that the supercurrent diode effect reported here does not rely upon subtle interfacial effects, which may play an increasing role when film thickness is comparable to the coherence length. An added benefit of using thicker films is that at 65 nm the Nb has a near bulk $T_c$, which makes our devices useful for integrating with Nb based computing schemes operating at 4.2 K.

Our data shows that the induced supercurrent diode effect depends upon the width of the track. From our data, we can estimate the upper limit of track width for the supercurrent diode effect. For the polycrystalline film, extrapolating the linear fit shown in Figures \ref{Fig2} (e) and (f) $\eta = 0$ for $w$ between about 12 and $13~\mu$m. In the epitaxial film, $\eta = 0$ is observed at 5 K for  $w = 7 \mu$m. Extrapolating the data at 1.8 K suggests that $\eta = 0$ for  $w \approx 8~\mu$m. In the epitaxial sample, $\eta$ also reduces in narrowest tracks, but this trend is not reproduced in the polycrystalline sample.

We propose that the induced diode effect does result from asymmetric properties in the two edges of our tracks. This asymmetry causes unequal generation, penetration and expulsion of vortices on the two edges of the tracks, resulting in the observed rectification. Each edge has a different critical current density required for penetration of magnetic vortices, $j_\textit{pen}$. This could be induced by the asymmetric edge roughness mechanism of Hou \textit{et al.} \cite{hou2022ubiquitous}, or more generally by any local changes in the superconducting gap such as, most probably, edge profile. The diode effect in our tracks therefore relies on being in a regime where the critical current density ($j_c$) of the tracks is limited by flux vortice entry. We know this to be the case from our analysis of the $I_c(B)$ dependence of Figure \ref{Fig1} (c), inset. The requirement for track to go normal is therefore flux penetrating the track from the edge barriers continuously and flux vortices moving freely across the track without being pinned.

There are three possible mechanisms to limit the critical current desnity $j_c$ of our superconducting tracks. The ultimate limit is the depairing current, $j_\textit{pair}$. Where the critical current of the track is limited by flux vortices, the critical current density is limited by the greater of the current density required for penetration of magnetic vortices, $j_\textit{pen}$, and the current density required to depin the vortices, $j_\textit{pin}$.

\begin{figure}
\includegraphics[width=0.5\columnwidth]{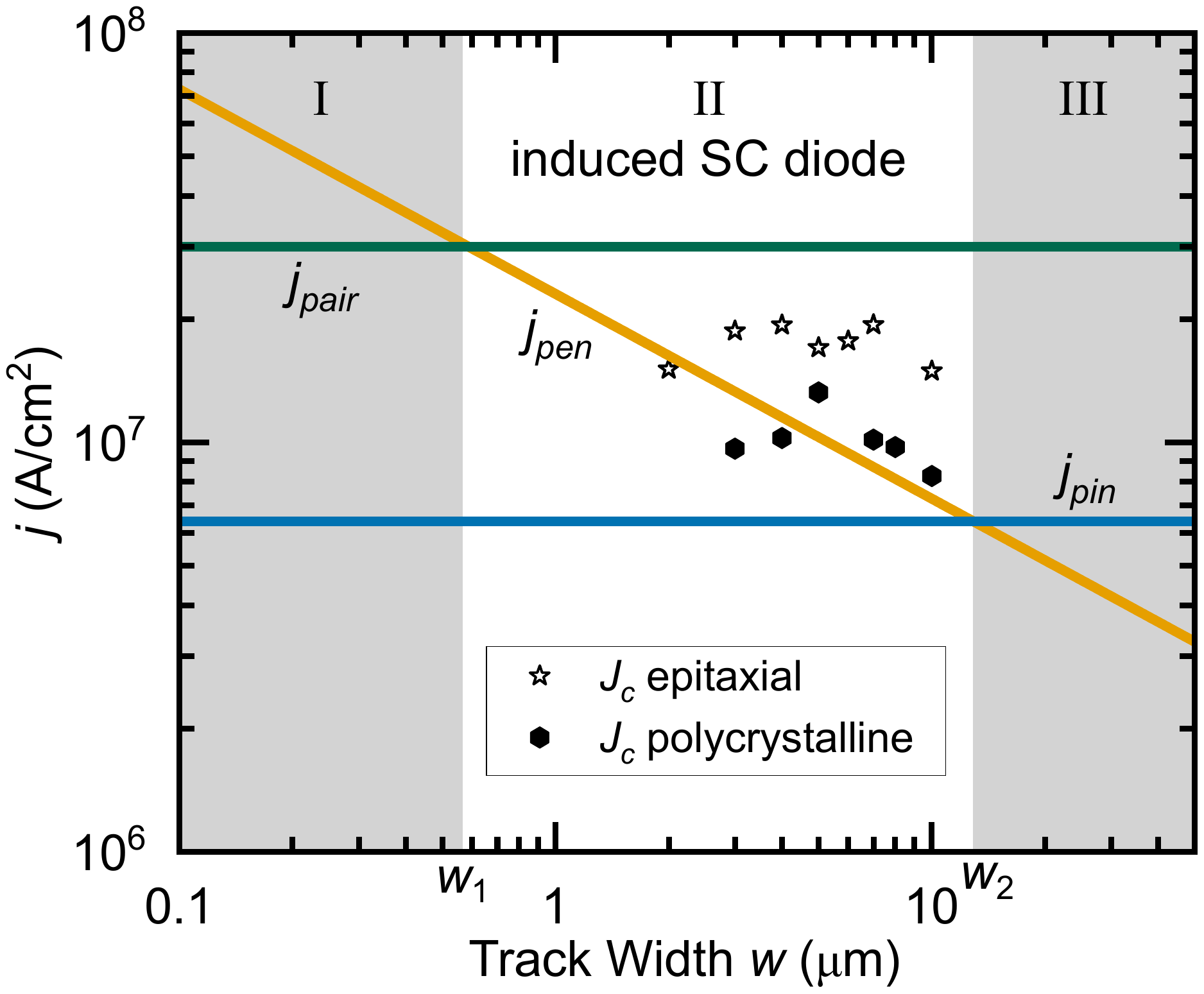}
\caption{\label{Fig8} Width dependence of the critical current densities in Nb tracks. Experimental data for the nominal critical current density at 5~K for all devices in this study. The polycrystalline tracks are fit to Equation \ref{jpen}, with a best fit value $B_\textit{pen} =11\pm1$~mT. Also shown are literature values for polycrystalline Nb depairing current\cite{ILIN2010953} ($j_\textit{pair}$) and depinning current\cite{ZHU19941357} ($j_\textit{pin}$). We define three regions depending on which mechanism determines the critical current density of our tracks, see text.}
\end{figure}

For our Nb tracks, we can reasonably assume that $j_\textit{pair}$ is an intrinsic material property and $j_\textit{pin}$ is a property of the film, both of which do not depend on track width. On the other hand, it has been shown that $j_\textit{pen}$ depends on the track width ($w$) according to\cite{ILIN2010953}
\begin{equation}
j_\textit{pen} = \frac{2\pi B_\textit{pen}}{\mu_0 \sqrt{dw}},
\label{jpen}
\end{equation}
\noindent where $d$ is the film thickness, $B_\textit{pen}$ is the minimum field for vortex penetration and $\mu_0$ is the magnetic permeability of free space. Figure \ref{Fig8} shows the experimentally determined maximum nominal $j_c$ at 5~K for both the epitaxial and polycrystalline Nb tracks. $j_\textit{pen}$ is fitted to the polycrystalline dataset with $B_\textit{pen}$ a free fitting parameter, returning a best fit value of $11\pm1$~mT. Note that $B_\textit{pen}$ is likely a combination of the applied and self magnetic fields in our system. Also plotted are the experimentally determined $j_\textit{pair}$ from Il’in \textit{et al.} for narrow polycrystalline Nb tracks\cite{ILIN2010953} and $j_\textit{pin}$ from Zhu \textit{et al.} for a 110~nm polycrystalline Nb film\cite{ZHU19941357}.

Considering the fit to the polycrystalline Nb tracks, it is possible to divide Figure \ref{Fig8} into three width regions. For narrow tracks occupying region I ($w<w_1$), the critical current density in the track is limited by $j_\textit{pair}$. Since vortices do not contribute to $j_c$, our conjecture is that it is not possible for fabrication to induce a supercurrent diode effect in this region. For intermediate region II ($w_1<w<w_2$), the critical current is limited by $j_\textit{pen}$. In this region, the track goes normal when the transport current equals $j_\textit{pen}$, and a diode effect can be induced by device fabrication if $j_\textit{pen}$ is asymmetric on each edges. For wide tracks in region III ($w>w_2$), the critical current is no longer limited by $j_\textit{pen}$. Here, upon applying a transport current, first vortices enter the track at the edges at transport current exceeding $j_\textit{pen}$. However, these vortices become pinned, preventing the track from going normal. It is only when the transport current exceeds $j_\textit{pin}$ that vortices can move and the track goes normal. In region III, the induced diode effect should be suppressed as $j_\textit{pin}$ is an intrinsic material property and therefore likely to be uniform across the width of the track.

Comparing the width dependence of $j_c$ in Figure \ref{Fig8} to our experimental data on the width dependence of the supercurrent diode effect in Figures \ref{Fig2} we can make the following observations and predictions. We expect that the induced diode effect will be suppressed in region III, which for the polycrystalline dataset is at $w\approx13~\mu$m. This width corresponds well with our extrapolation to $\eta = 0$ in Figures \ref{Fig2} (e) and (f). We can also use the model to predict that $w=600$ nm is the upper boundary of region I for polycrystalline Nb. Therefore tracks narrower than about 600 nm should not show induced supercurrent diode effect, however fabrication of tracks that narrow would require a change in fabrication methodology. Finally, applying similar analysis to the epitaxial Nb tracks, we conclude that both $j_\textit{pen}$ and $j_\textit{pin}$ must be higher in the epitaxial Nb. Interestingly, a higher $j_\textit{pen}$ would suggest that that region I extends to wider epitaxial tracks tracks compared to polycrystalline tracks, which would be consistent with our observation of reduced $\eta$ in the narrowest epitaxial tracks in Figures \ref{Fig2} (b) and (c).

\section{Conclusions}

In conclusion, we report on induced supercurrent diode effect in 65 nm thick polycrystalline and epitaxial Nb films patterned into tracks. Consistent with previous works, the superconductor itself does not require any intrinsic inversion symmetry breaking and the layer can be significantly thicker than the coherence length. Our experimental results and model suggests that in order to distinguish between intrinsic and induced supercurrent diode effect in a materials system, the critical current density in the fabricated devices should be limited by $j_\textit{pair}$ and not by flux vortices. Therefore, claims of intrinisic diode effect should only be made for devices limited by $j_\textit{pair}$ (occupying region I of Figure \ref{Fig8}). This is particularly relevant for the study of exotic materials to isolate the proposed intrinsic effect from the fabrication induced effect we report here.

For technological applications, it would be possible to manipulate the devices edges to create local differences in $j_\textit{pen}$, for example by engineering edge roughness or by changing the superconducting properties of one edge by ion implantation, and therefore control induced diode effects.

 Our largest reported diode efficency is $\eta \approx30\%$ in a 4~$\mu$m wide track at 5 K. We report that the supercurrent diode effect can be observed in Nb tracks over a range of track widths, applied out-of-plane applied fields and temperatures. Our results show a track width dependence, where for the widest tracks in this study, we report that the diode effect reduces or disappears altogether. We present a model of the critical currents in the track to account for our observed width dependence.

\begin{acknowledgments}
The work was supported financially through the following EPSRC grant: EP/V028138/1. We acknowledge support from the Henry Royce Institute. We would like to thank N.O. Birge for helpful discussions. 
\end{acknowledgments}

\section*{AUTHOR DECLARATIONS}

\subsection*{Conflict of Interest}
The authors have no conflicts to disclose.

\subsection*{Author Contributions}

\noindent\textbf{N Satchell}: Investigation (lead); methodology (lead); conceptualization (equal); writing - original draft (lead); visualization (lead); formal analysis (lead); writing - review and editing (equal); funding acquisition (supporting).
\textbf{PM Shepley}: Investigation (supporting); methodology (supporting); writing - review and editing (equal).
\textbf{MC Rosamond}: Investigation (supporting); methodology (supporting); writing - review and editing (equal).
\textbf{G Burnell}: Project administration (lead); funding acquisition (lead); conceptualization (equal); methodology (supporting); investigation (supporting); writing - review and editing (equal).

\section*{DATA AVAILABILITY}

The datasets generated during the current study are available in the University of Leeds repository, DOI: \href{https://doi.org/10.5518/1293}{10.5518/1293}.


\bibliography{diodes}

\begin{thebibliography}{57}%
\makeatletter
\providecommand \@ifxundefined [1]{%
 \@ifx{#1\undefined}
}%
\providecommand \@ifnum [1]{%
 \ifnum #1\expandafter \@firstoftwo
 \else \expandafter \@secondoftwo
 \fi
}%
\providecommand \@ifx [1]{%
 \ifx #1\expandafter \@firstoftwo
 \else \expandafter \@secondoftwo
 \fi
}%
\providecommand \natexlab [1]{#1}%
\providecommand \enquote  [1]{``#1''}%
\providecommand \bibnamefont  [1]{#1}%
\providecommand \bibfnamefont [1]{#1}%
\providecommand \citenamefont [1]{#1}%
\providecommand \href@noop [0]{\@secondoftwo}%
\providecommand \href [0]{\begingroup \@sanitize@url \@href}%
\providecommand \@href[1]{\@@startlink{#1}\@@href}%
\providecommand \@@href[1]{\endgroup#1\@@endlink}%
\providecommand \@sanitize@url [0]{\catcode `\\12\catcode `\$12\catcode
  `\&12\catcode `\#12\catcode `\^12\catcode `\_12\catcode `\%12\relax}%
\providecommand \@@startlink[1]{}%
\providecommand \@@endlink[0]{}%
\providecommand \url  [0]{\begingroup\@sanitize@url \@url }%
\providecommand \@url [1]{\endgroup\@href {#1}{\urlprefix }}%
\providecommand \urlprefix  [0]{URL }%
\providecommand \Eprint [0]{\href }%
\providecommand \doibase [0]{http://dx.doi.org/}%
\providecommand \selectlanguage [0]{\@gobble}%
\providecommand \bibinfo  [0]{\@secondoftwo}%
\providecommand \bibfield  [0]{\@secondoftwo}%
\providecommand \translation [1]{[#1]}%
\providecommand \BibitemOpen [0]{}%
\providecommand \bibitemStop [0]{}%
\providecommand \bibitemNoStop [0]{.\EOS\space}%
\providecommand \EOS [0]{\spacefactor3000\relax}%
\providecommand \BibitemShut  [1]{\csname bibitem#1\endcsname}%
\let\auto@bib@innerbib\@empty
\bibitem [{\citenamefont {Swartz}\ and\ \citenamefont
  {Hart}(1967)}]{PhysRev.156.412}%
  \BibitemOpen
  \bibfield  {author} {\bibinfo {author} {\bibfnamefont {P.~S.}\ \bibnamefont
  {Swartz}}\ and\ \bibinfo {author} {\bibfnamefont {H.~R.}\ \bibnamefont
  {Hart}},\ }\bibfield  {title} {\enquote {\bibinfo {title} {Asymmetries of the
  {C}ritical {S}urface {C}urrent in {T}ype-{II} {S}uperconductors},}\ }\href
  {\doibase 10.1103/PhysRev.156.412} {\bibfield  {journal} {\bibinfo  {journal}
  {Phys. Rev.}\ }\textbf {\bibinfo {volume} {156}},\ \bibinfo {pages}
  {412--420} (\bibinfo {year} {1967})}\BibitemShut {NoStop}%
\bibitem [{\citenamefont {Touitou}\ \emph {et~al.}(2004)\citenamefont
  {Touitou}, \citenamefont {Bernstein}, \citenamefont {Hamet}, \citenamefont
  {Simon}, \citenamefont {Méchin}, \citenamefont {Contour},\ and\
  \citenamefont {Jacquet}}]{doi:10.1063/1.1789231}%
  \BibitemOpen
  \bibfield  {author} {\bibinfo {author} {\bibfnamefont {N.}~\bibnamefont
  {Touitou}}, \bibinfo {author} {\bibfnamefont {P.}~\bibnamefont {Bernstein}},
  \bibinfo {author} {\bibfnamefont {J.~F.}\ \bibnamefont {Hamet}}, \bibinfo
  {author} {\bibfnamefont {C.}~\bibnamefont {Simon}}, \bibinfo {author}
  {\bibfnamefont {L.}~\bibnamefont {Méchin}}, \bibinfo {author} {\bibfnamefont
  {J.~P.}\ \bibnamefont {Contour}}, \ and\ \bibinfo {author} {\bibfnamefont
  {E.}~\bibnamefont {Jacquet}},\ }\bibfield  {title} {\enquote {\bibinfo
  {title} {Nonsymmetric current–voltage characteristics in
  ferromagnet/superconductor thin film structures},}\ }\href {\doibase
  10.1063/1.1789231} {\bibfield  {journal} {\bibinfo  {journal} {Appl. Phys.
  Lett.}\ }\textbf {\bibinfo {volume} {85}},\ \bibinfo {pages} {1742--1744}
  (\bibinfo {year} {2004})}\BibitemShut {NoStop}%
\bibitem [{\citenamefont {Vodolazov}\ \emph {et~al.}(2005)\citenamefont
  {Vodolazov}, \citenamefont {Gribkov}, \citenamefont {Gusev}, \citenamefont
  {Klimov}, \citenamefont {Nozdrin}, \citenamefont {Rogov},\ and\ \citenamefont
  {Vdovichev}}]{PhysRevB.72.064509}%
  \BibitemOpen
  \bibfield  {author} {\bibinfo {author} {\bibfnamefont {D.~Y.}\ \bibnamefont
  {Vodolazov}}, \bibinfo {author} {\bibfnamefont {B.~A.}\ \bibnamefont
  {Gribkov}}, \bibinfo {author} {\bibfnamefont {S.~A.}\ \bibnamefont {Gusev}},
  \bibinfo {author} {\bibfnamefont {A.~Y.}\ \bibnamefont {Klimov}}, \bibinfo
  {author} {\bibfnamefont {Y.~N.}\ \bibnamefont {Nozdrin}}, \bibinfo {author}
  {\bibfnamefont {V.~V.}\ \bibnamefont {Rogov}}, \ and\ \bibinfo {author}
  {\bibfnamefont {S.~N.}\ \bibnamefont {Vdovichev}},\ }\bibfield  {title}
  {\enquote {\bibinfo {title} {Considerable enhancement of the critical current
  in a superconducting film by a magnetized magnetic strip},}\ }\href {\doibase
  10.1103/PhysRevB.72.064509} {\bibfield  {journal} {\bibinfo  {journal} {Phys.
  Rev. B}\ }\textbf {\bibinfo {volume} {72}},\ \bibinfo {pages} {064509}
  (\bibinfo {year} {2005})}\BibitemShut {NoStop}%
\bibitem [{\citenamefont {Morelle}\ and\ \citenamefont
  {Moshchalkov}(2006)}]{doi:10.1063/1.2199468}%
  \BibitemOpen
  \bibfield  {author} {\bibinfo {author} {\bibfnamefont {M.}~\bibnamefont
  {Morelle}}\ and\ \bibinfo {author} {\bibfnamefont {V.~V.}\ \bibnamefont
  {Moshchalkov}},\ }\bibfield  {title} {\enquote {\bibinfo {title} {Enhanced
  critical currents through field compensation with magnetic strips},}\ }\href
  {\doibase 10.1063/1.2199468} {\bibfield  {journal} {\bibinfo  {journal}
  {Appl. Phys. Lett.}\ }\textbf {\bibinfo {volume} {88}},\ \bibinfo {pages}
  {172507} (\bibinfo {year} {2006})}\BibitemShut {NoStop}%
\bibitem [{\citenamefont {Papon}, \citenamefont {Senapati},\ and\ \citenamefont
  {Barber}(2008)}]{doi:10.1063/1.3009207}%
  \BibitemOpen
  \bibfield  {author} {\bibinfo {author} {\bibfnamefont {A.}~\bibnamefont
  {Papon}}, \bibinfo {author} {\bibfnamefont {K.}~\bibnamefont {Senapati}}, \
  and\ \bibinfo {author} {\bibfnamefont {Z.~H.}\ \bibnamefont {Barber}},\
  }\bibfield  {title} {\enquote {\bibinfo {title} {Asymmetric critical current
  of niobium microbridges with ferromagnetic stripe},}\ }\href {\doibase
  10.1063/1.3009207} {\bibfield  {journal} {\bibinfo  {journal} {Appl. Phys.
  Lett.}\ }\textbf {\bibinfo {volume} {93}},\ \bibinfo {pages} {172507}
  (\bibinfo {year} {2008})}\BibitemShut {NoStop}%
\bibitem [{\citenamefont {Grein}\ \emph {et~al.}(2009)\citenamefont {Grein},
  \citenamefont {Eschrig}, \citenamefont {Metalidis},\ and\ \citenamefont
  {Sch\"on}}]{PhysRevLett.102.227005}%
  \BibitemOpen
  \bibfield  {author} {\bibinfo {author} {\bibfnamefont {R.}~\bibnamefont
  {Grein}}, \bibinfo {author} {\bibfnamefont {M.}~\bibnamefont {Eschrig}},
  \bibinfo {author} {\bibfnamefont {G.}~\bibnamefont {Metalidis}}, \ and\
  \bibinfo {author} {\bibfnamefont {G.}~\bibnamefont {Sch\"on}},\ }\bibfield
  {title} {\enquote {\bibinfo {title} {Spin-{Dependent Cooper Pair Phase and
  Pure Spin Supercurrents in Strongly Polarized F}erromagnets},}\ }\href
  {\doibase 10.1103/PhysRevLett.102.227005} {\bibfield  {journal} {\bibinfo
  {journal} {Phys. Rev. Lett.}\ }\textbf {\bibinfo {volume} {102}},\ \bibinfo
  {pages} {227005} (\bibinfo {year} {2009})}\BibitemShut {NoStop}%
\bibitem [{\citenamefont {Halterman}\ \emph {et~al.}(2022)\citenamefont
  {Halterman}, \citenamefont {Alidoust}, \citenamefont {Smith},\ and\
  \citenamefont {Starr}}]{PhysRevB.105.104508}%
  \BibitemOpen
  \bibfield  {author} {\bibinfo {author} {\bibfnamefont {K.}~\bibnamefont
  {Halterman}}, \bibinfo {author} {\bibfnamefont {M.}~\bibnamefont {Alidoust}},
  \bibinfo {author} {\bibfnamefont {R.}~\bibnamefont {Smith}}, \ and\ \bibinfo
  {author} {\bibfnamefont {S.}~\bibnamefont {Starr}},\ }\bibfield  {title}
  {\enquote {\bibinfo {title} {Supercurrent diode effect, spin torques, and
  robust zero-energy peak in planar half-metallic trilayers},}\ }\href
  {\doibase 10.1103/PhysRevB.105.104508} {\bibfield  {journal} {\bibinfo
  {journal} {Phys. Rev. B}\ }\textbf {\bibinfo {volume} {105}},\ \bibinfo
  {pages} {104508} (\bibinfo {year} {2022})}\BibitemShut {NoStop}%
\bibitem [{\citenamefont {Zhang}\ \emph {et~al.}(2022)\citenamefont {Zhang},
  \citenamefont {Gu}, \citenamefont {Li}, \citenamefont {Hu},\ and\
  \citenamefont {Jiang}}]{PhysRevX.12.041013}%
  \BibitemOpen
  \bibfield  {author} {\bibinfo {author} {\bibfnamefont {Y.}~\bibnamefont
  {Zhang}}, \bibinfo {author} {\bibfnamefont {Y.}~\bibnamefont {Gu}}, \bibinfo
  {author} {\bibfnamefont {P.}~\bibnamefont {Li}}, \bibinfo {author}
  {\bibfnamefont {J.}~\bibnamefont {Hu}}, \ and\ \bibinfo {author}
  {\bibfnamefont {K.}~\bibnamefont {Jiang}},\ }\bibfield  {title} {\enquote
  {\bibinfo {title} {General {T}heory of {J}osephson {D}iodes},}\ }\href
  {\doibase 10.1103/PhysRevX.12.041013} {\bibfield  {journal} {\bibinfo
  {journal} {Phys. Rev. X}\ }\textbf {\bibinfo {volume} {12}},\ \bibinfo
  {pages} {041013} (\bibinfo {year} {2022})}\BibitemShut {NoStop}%
\bibitem [{\citenamefont {Davydova}, \citenamefont {Prembabu},\ and\
  \citenamefont {Fu}(2022)}]{davydova2022universal}%
  \BibitemOpen
  \bibfield  {author} {\bibinfo {author} {\bibfnamefont {M.}~\bibnamefont
  {Davydova}}, \bibinfo {author} {\bibfnamefont {S.}~\bibnamefont {Prembabu}},
  \ and\ \bibinfo {author} {\bibfnamefont {L.}~\bibnamefont {Fu}},\ }\bibfield
  {title} {\enquote {\bibinfo {title} {Universal {J}osephson diode effect},}\
  }\href {\doibase https://doi.org/10.1126/sciadv.abo0309} {\bibfield
  {journal} {\bibinfo  {journal} {Sci. Adv.}\ }\textbf {\bibinfo {volume}
  {8}},\ \bibinfo {pages} {eabo0309} (\bibinfo {year} {2022})}\BibitemShut
  {NoStop}%
\bibitem [{\citenamefont {Eschrig}(2015)}]{Eschrig_2015}%
  \BibitemOpen
  \bibfield  {author} {\bibinfo {author} {\bibfnamefont {M.}~\bibnamefont
  {Eschrig}},\ }\bibfield  {title} {\enquote {\bibinfo {title} {Spin-polarized
  supercurrents for spintronics: a review of current progress},}\ }\href
  {\doibase 10.1088/0034-4885/78/10/104501} {\bibfield  {journal} {\bibinfo
  {journal} {Rep. Prog. Phys.}\ }\textbf {\bibinfo {volume} {78}},\ \bibinfo
  {pages} {104501} (\bibinfo {year} {2015})}\BibitemShut {NoStop}%
\bibitem [{\citenamefont {Braude}\ and\ \citenamefont
  {Nazarov}(2007)}]{PhysRevLett.98.077003}%
  \BibitemOpen
  \bibfield  {author} {\bibinfo {author} {\bibfnamefont {V.}~\bibnamefont
  {Braude}}\ and\ \bibinfo {author} {\bibfnamefont {Y.~V.}\ \bibnamefont
  {Nazarov}},\ }\bibfield  {title} {\enquote {\bibinfo {title} {Fully
  {D}eveloped {T}riplet {P}roximity {E}ffect},}\ }\href {\doibase
  10.1103/PhysRevLett.98.077003} {\bibfield  {journal} {\bibinfo  {journal}
  {Phys. Rev. Lett.}\ }\textbf {\bibinfo {volume} {98}},\ \bibinfo {pages}
  {077003} (\bibinfo {year} {2007})}\BibitemShut {NoStop}%
\bibitem [{\citenamefont {Heim}\ \emph {et~al.}(2013)\citenamefont {Heim},
  \citenamefont {Pugach}, \citenamefont {Kupriyanov}, \citenamefont {Goldobin},
  \citenamefont {Koelle},\ and\ \citenamefont {Kleiner}}]{Heim_2013}%
  \BibitemOpen
  \bibfield  {author} {\bibinfo {author} {\bibfnamefont {D.~M.}\ \bibnamefont
  {Heim}}, \bibinfo {author} {\bibfnamefont {N.~G.}\ \bibnamefont {Pugach}},
  \bibinfo {author} {\bibfnamefont {M.~Y.}\ \bibnamefont {Kupriyanov}},
  \bibinfo {author} {\bibfnamefont {E.}~\bibnamefont {Goldobin}}, \bibinfo
  {author} {\bibfnamefont {D.}~\bibnamefont {Koelle}}, \ and\ \bibinfo {author}
  {\bibfnamefont {R.}~\bibnamefont {Kleiner}},\ }\bibfield  {title} {\enquote
  {\bibinfo {title} {Ferromagnetic planar josephson junction with transparent
  interfaces: a $\phi$ junction proposal},}\ }\href {\doibase
  10.1088/0953-8984/25/21/215701} {\bibfield  {journal} {\bibinfo  {journal}
  {J. Phys. Condens. Matter}\ }\textbf {\bibinfo {volume} {25}},\ \bibinfo
  {pages} {215701} (\bibinfo {year} {2013})}\BibitemShut {NoStop}%
\bibitem [{\citenamefont {Goldobin}, \citenamefont {Koelle},\ and\
  \citenamefont {Kleiner}(2015)}]{PhysRevB.91.214511}%
  \BibitemOpen
  \bibfield  {author} {\bibinfo {author} {\bibfnamefont {E.}~\bibnamefont
  {Goldobin}}, \bibinfo {author} {\bibfnamefont {D.}~\bibnamefont {Koelle}}, \
  and\ \bibinfo {author} {\bibfnamefont {R.}~\bibnamefont {Kleiner}},\
  }\bibfield  {title} {\enquote {\bibinfo {title} {Tunable
  $\ifmmode\pm\else\textpm\fi{}\ensuremath{\varphi},\phantom{\rule{0.16em}{0ex}}{\ensuremath{\varphi}}_{0}$,
  and
  ${\ensuremath{\varphi}}_{0}\ifmmode\pm\else\textpm\fi{}\ensuremath{\varphi}$
  josephson junction},}\ }\href {\doibase 10.1103/PhysRevB.91.214511}
  {\bibfield  {journal} {\bibinfo  {journal} {Phys. Rev. B}\ }\textbf {\bibinfo
  {volume} {91}},\ \bibinfo {pages} {214511} (\bibinfo {year}
  {2015})}\BibitemShut {NoStop}%
\bibitem [{\citenamefont {Shukrinov}, \citenamefont {Rahmonov},\ and\
  \citenamefont {Sengupta}(2019)}]{PhysRevB.99.224513}%
  \BibitemOpen
  \bibfield  {author} {\bibinfo {author} {\bibfnamefont {Y.~M.}\ \bibnamefont
  {Shukrinov}}, \bibinfo {author} {\bibfnamefont {I.~R.}\ \bibnamefont
  {Rahmonov}}, \ and\ \bibinfo {author} {\bibfnamefont {K.}~\bibnamefont
  {Sengupta}},\ }\bibfield  {title} {\enquote {\bibinfo {title} {Ferromagnetic
  resonance and magnetic precessions in ${\ensuremath{\varphi}}_{0}$
  junctions},}\ }\href {\doibase 10.1103/PhysRevB.99.224513} {\bibfield
  {journal} {\bibinfo  {journal} {Phys. Rev. B}\ }\textbf {\bibinfo {volume}
  {99}},\ \bibinfo {pages} {224513} (\bibinfo {year} {2019})}\BibitemShut
  {NoStop}%
\bibitem [{\citenamefont {Mazanik}\ \emph {et~al.}(2020)\citenamefont
  {Mazanik}, \citenamefont {Rahmonov}, \citenamefont {Botha},\ and\
  \citenamefont {Shukrinov}}]{PhysRevApplied.14.014003}%
  \BibitemOpen
  \bibfield  {author} {\bibinfo {author} {\bibfnamefont {A.}~\bibnamefont
  {Mazanik}}, \bibinfo {author} {\bibfnamefont {I.}~\bibnamefont {Rahmonov}},
  \bibinfo {author} {\bibfnamefont {A.}~\bibnamefont {Botha}}, \ and\ \bibinfo
  {author} {\bibfnamefont {Y.}~\bibnamefont {Shukrinov}},\ }\bibfield  {title}
  {\enquote {\bibinfo {title} {Analytical criteria for magnetization reversal
  in a ${\ensuremath{\varphi}}_{0}$ josephson junction},}\ }\href {\doibase
  10.1103/PhysRevApplied.14.014003} {\bibfield  {journal} {\bibinfo  {journal}
  {Phys. Rev. Appl.}\ }\textbf {\bibinfo {volume} {14}},\ \bibinfo {pages}
  {014003} (\bibinfo {year} {2020})}\BibitemShut {NoStop}%
\bibitem [{\citenamefont {Meng}\ \emph {et~al.}(2022)\citenamefont {Meng},
  \citenamefont {Wu}, \citenamefont {Ren},\ and\ \citenamefont
  {Wu}}]{PhysRevB.106.174502}%
  \BibitemOpen
  \bibfield  {author} {\bibinfo {author} {\bibfnamefont {H.}~\bibnamefont
  {Meng}}, \bibinfo {author} {\bibfnamefont {X.}~\bibnamefont {Wu}}, \bibinfo
  {author} {\bibfnamefont {Y.}~\bibnamefont {Ren}}, \ and\ \bibinfo {author}
  {\bibfnamefont {J.}~\bibnamefont {Wu}},\ }\bibfield  {title} {\enquote
  {\bibinfo {title} {Anomalous supercurrent modulated by interfacial
  magnetizations in {J}osephson junctions with ferromagnetic bilayers},}\
  }\href {\doibase 10.1103/PhysRevB.106.174502} {\bibfield  {journal} {\bibinfo
   {journal} {Phys. Rev. B}\ }\textbf {\bibinfo {volume} {106}},\ \bibinfo
  {pages} {174502} (\bibinfo {year} {2022})}\BibitemShut {NoStop}%
\bibitem [{\citenamefont {Bobkova}\ \emph {et~al.}(2016)\citenamefont
  {Bobkova}, \citenamefont {Bobkov}, \citenamefont {Zyuzin},\ and\
  \citenamefont {Alidoust}}]{PhysRevB.94.134506}%
  \BibitemOpen
  \bibfield  {author} {\bibinfo {author} {\bibfnamefont {I.~V.}\ \bibnamefont
  {Bobkova}}, \bibinfo {author} {\bibfnamefont {A.~M.}\ \bibnamefont {Bobkov}},
  \bibinfo {author} {\bibfnamefont {A.~A.}\ \bibnamefont {Zyuzin}}, \ and\
  \bibinfo {author} {\bibfnamefont {M.}~\bibnamefont {Alidoust}},\ }\bibfield
  {title} {\enquote {\bibinfo {title} {Magnetoelectrics in disordered
  topological insulator {J}osephson junctions},}\ }\href {\doibase
  10.1103/PhysRevB.94.134506} {\bibfield  {journal} {\bibinfo  {journal} {Phys.
  Rev. B}\ }\textbf {\bibinfo {volume} {94}},\ \bibinfo {pages} {134506}
  (\bibinfo {year} {2016})}\BibitemShut {NoStop}%
\bibitem [{\citenamefont {Alidoust}\ and\ \citenamefont
  {Hamzehpour}(2017)}]{PhysRevB.96.165422}%
  \BibitemOpen
  \bibfield  {author} {\bibinfo {author} {\bibfnamefont {M.}~\bibnamefont
  {Alidoust}}\ and\ \bibinfo {author} {\bibfnamefont {H.}~\bibnamefont
  {Hamzehpour}},\ }\bibfield  {title} {\enquote {\bibinfo {title} {Spontaneous
  supercurrent and ${\ensuremath{\varphi}}_{0}$ phase shift parallel to
  magnetized topological insulator interfaces},}\ }\href {\doibase
  10.1103/PhysRevB.96.165422} {\bibfield  {journal} {\bibinfo  {journal} {Phys.
  Rev. B}\ }\textbf {\bibinfo {volume} {96}},\ \bibinfo {pages} {165422}
  (\bibinfo {year} {2017})}\BibitemShut {NoStop}%
\bibitem [{\citenamefont {Li}\ \emph {et~al.}(2022)\citenamefont {Li},
  \citenamefont {Yang}, \citenamefont {Hu}, \citenamefont {Liu},\ and\
  \citenamefont {Tao}}]{Li_2022}%
  \BibitemOpen
  \bibfield  {author} {\bibinfo {author} {\bibfnamefont {M.~Y.}\ \bibnamefont
  {Li}}, \bibinfo {author} {\bibfnamefont {Y.}~\bibnamefont {Yang}}, \bibinfo
  {author} {\bibfnamefont {J.~G.}\ \bibnamefont {Hu}}, \bibinfo {author}
  {\bibfnamefont {T.~M.}\ \bibnamefont {Liu}}, \ and\ \bibinfo {author}
  {\bibfnamefont {Y.~C.}\ \bibnamefont {Tao}},\ }\bibfield  {title} {\enquote
  {\bibinfo {title} {Anomalous {J}osephson current through a topological
  noncoplanar ferromagnetic trilayer},}\ }\href {\doibase
  10.1088/1361-648X/ac484b} {\bibfield  {journal} {\bibinfo  {journal} {J.
  Phys. Condens. Matter}\ }\textbf {\bibinfo {volume} {34}},\ \bibinfo {pages}
  {135801} (\bibinfo {year} {2022})}\BibitemShut {NoStop}%
\bibitem [{\citenamefont {Alidoust}(2018)}]{PhysRevB.98.245418}%
  \BibitemOpen
  \bibfield  {author} {\bibinfo {author} {\bibfnamefont {M.}~\bibnamefont
  {Alidoust}},\ }\bibfield  {title} {\enquote {\bibinfo {title} {Self-biased
  current, magnetic interference response, and superconducting vortices in
  tilted {W}eyl semimetals with disorder},}\ }\href {\doibase
  10.1103/PhysRevB.98.245418} {\bibfield  {journal} {\bibinfo  {journal} {Phys.
  Rev. B}\ }\textbf {\bibinfo {volume} {98}},\ \bibinfo {pages} {245418}
  (\bibinfo {year} {2018})}\BibitemShut {NoStop}%
\bibitem [{\citenamefont {Alidoust}\ and\ \citenamefont
  {Halterman}(2020)}]{PhysRevB.101.035120}%
  \BibitemOpen
  \bibfield  {author} {\bibinfo {author} {\bibfnamefont {M.}~\bibnamefont
  {Alidoust}}\ and\ \bibinfo {author} {\bibfnamefont {K.}~\bibnamefont
  {Halterman}},\ }\bibfield  {title} {\enquote {\bibinfo {title} {Evolution of
  pair correlation symmetries and supercurrent reversal in tilted {W}eyl
  semimetals},}\ }\href {\doibase 10.1103/PhysRevB.101.035120} {\bibfield
  {journal} {\bibinfo  {journal} {Phys. Rev. B}\ }\textbf {\bibinfo {volume}
  {101}},\ \bibinfo {pages} {035120} (\bibinfo {year} {2020})}\BibitemShut
  {NoStop}%
\bibitem [{\citenamefont {Kulikov}\ \emph {et~al.}(2020)\citenamefont
  {Kulikov}, \citenamefont {Sinha}, \citenamefont {Shukrinov},\ and\
  \citenamefont {Sengupta}}]{PhysRevB.101.075110}%
  \BibitemOpen
  \bibfield  {author} {\bibinfo {author} {\bibfnamefont {K.}~\bibnamefont
  {Kulikov}}, \bibinfo {author} {\bibfnamefont {D.}~\bibnamefont {Sinha}},
  \bibinfo {author} {\bibfnamefont {Y.~M.}\ \bibnamefont {Shukrinov}}, \ and\
  \bibinfo {author} {\bibfnamefont {K.}~\bibnamefont {Sengupta}},\ }\bibfield
  {title} {\enquote {\bibinfo {title} {Josephson junctions of {W}eyl and
  multi-{W}eyl semimetals},}\ }\href {\doibase 10.1103/PhysRevB.101.075110}
  {\bibfield  {journal} {\bibinfo  {journal} {Phys. Rev. B}\ }\textbf {\bibinfo
  {volume} {101}},\ \bibinfo {pages} {075110} (\bibinfo {year}
  {2020})}\BibitemShut {NoStop}%
\bibitem [{\citenamefont {Sinha}(2020)}]{PhysRevB.102.085144}%
  \BibitemOpen
  \bibfield  {author} {\bibinfo {author} {\bibfnamefont {D.}~\bibnamefont
  {Sinha}},\ }\bibfield  {title} {\enquote {\bibinfo {title} {Josephson effect
  in type-{I W}eyl semimetals},}\ }\href {\doibase 10.1103/PhysRevB.102.085144}
  {\bibfield  {journal} {\bibinfo  {journal} {Phys. Rev. B}\ }\textbf {\bibinfo
  {volume} {102}},\ \bibinfo {pages} {085144} (\bibinfo {year}
  {2020})}\BibitemShut {NoStop}%
\bibitem [{\citenamefont {Alidoust}(2020)}]{PhysRevB.101.155123}%
  \BibitemOpen
  \bibfield  {author} {\bibinfo {author} {\bibfnamefont {M.}~\bibnamefont
  {Alidoust}},\ }\bibfield  {title} {\enquote {\bibinfo {title} {Critical
  supercurrent and ${\ensuremath{\varphi}}_{0}$ state for probing a persistent
  spin helix},}\ }\href {\doibase 10.1103/PhysRevB.101.155123} {\bibfield
  {journal} {\bibinfo  {journal} {Phys. Rev. B}\ }\textbf {\bibinfo {volume}
  {101}},\ \bibinfo {pages} {155123} (\bibinfo {year} {2020})}\BibitemShut
  {NoStop}%
\bibitem [{\citenamefont {Alidoust}, \citenamefont {Shen},\ and\ \citenamefont
  {\ifmmode \check{Z}\else \v{Z}\fi{}uti\ifmmode~\acute{c}\else
  \'{c}\fi{}}(2021)}]{PhysRevB.103.L060503}%
  \BibitemOpen
  \bibfield  {author} {\bibinfo {author} {\bibfnamefont {M.}~\bibnamefont
  {Alidoust}}, \bibinfo {author} {\bibfnamefont {C.}~\bibnamefont {Shen}}, \
  and\ \bibinfo {author} {\bibfnamefont {I.}~\bibnamefont {\ifmmode
  \check{Z}\else \v{Z}\fi{}uti\ifmmode~\acute{c}\else \'{c}\fi{}}},\ }\bibfield
   {title} {\enquote {\bibinfo {title} {Cubic spin-orbit coupling and anomalous
  {J}osephson effect in planar junctions},}\ }\href {\doibase
  10.1103/PhysRevB.103.L060503} {\bibfield  {journal} {\bibinfo  {journal}
  {Phys. Rev. B}\ }\textbf {\bibinfo {volume} {103}},\ \bibinfo {pages}
  {L060503} (\bibinfo {year} {2021})}\BibitemShut {NoStop}%
\bibitem [{\citenamefont {Monroe}, \citenamefont {Alidoust},\ and\
  \citenamefont {\ifmmode \check{Z}\else \v{Z}\fi{}uti\ifmmode~\acute{c}\else
  \'{c}\fi{}}(2022)}]{PhysRevApplied.18.L031001}%
  \BibitemOpen
  \bibfield  {author} {\bibinfo {author} {\bibfnamefont {D.}~\bibnamefont
  {Monroe}}, \bibinfo {author} {\bibfnamefont {M.}~\bibnamefont {Alidoust}}, \
  and\ \bibinfo {author} {\bibfnamefont {I.}~\bibnamefont {\ifmmode
  \check{Z}\else \v{Z}\fi{}uti\ifmmode~\acute{c}\else \'{c}\fi{}}},\ }\bibfield
   {title} {\enquote {\bibinfo {title} {Tunable {Planar Josephson Junctions
  Driven by Time-Dependent Spin-Orbit C}oupling},}\ }\href {\doibase
  10.1103/PhysRevApplied.18.L031001} {\bibfield  {journal} {\bibinfo  {journal}
  {Phys. Rev. Appl.}\ }\textbf {\bibinfo {volume} {18}},\ \bibinfo {pages}
  {L031001} (\bibinfo {year} {2022})}\BibitemShut {NoStop}%
\bibitem [{\citenamefont {Amin}, \citenamefont {Omelyanchouk},\ and\
  \citenamefont {Zagoskin}(2001)}]{PhysRevB.63.212502}%
  \BibitemOpen
  \bibfield  {author} {\bibinfo {author} {\bibfnamefont {M.~H.~S.}\
  \bibnamefont {Amin}}, \bibinfo {author} {\bibfnamefont {A.~N.}\ \bibnamefont
  {Omelyanchouk}}, \ and\ \bibinfo {author} {\bibfnamefont {A.~M.}\
  \bibnamefont {Zagoskin}},\ }\bibfield  {title} {\enquote {\bibinfo {title}
  {Mechanisms of spontaneous current generation in an inhomogeneous d-wave
  superconductor},}\ }\href {\doibase 10.1103/PhysRevB.63.212502} {\bibfield
  {journal} {\bibinfo  {journal} {Phys. Rev. B}\ }\textbf {\bibinfo {volume}
  {63}},\ \bibinfo {pages} {212502} (\bibinfo {year} {2001})}\BibitemShut
  {NoStop}%
\bibitem [{\citenamefont {Liu}, \citenamefont {Zhou},\ and\ \citenamefont
  {Tao}(2021)}]{Liu_2021}%
  \BibitemOpen
  \bibfield  {author} {\bibinfo {author} {\bibfnamefont {T.}~\bibnamefont
  {Liu}}, \bibinfo {author} {\bibfnamefont {L.}~\bibnamefont {Zhou}}, \ and\
  \bibinfo {author} {\bibfnamefont {Y.~C.}\ \bibnamefont {Tao}},\ }\bibfield
  {title} {\enquote {\bibinfo {title} {Anomalous {J}osephson effect modulated
  by magnetic misorientation in a topological unconventional superconductor
  hybrid structure},}\ }\href {\doibase 10.1209/0295-5075/ac2f0e} {\bibfield
  {journal} {\bibinfo  {journal} {EPL}\ }\textbf {\bibinfo {volume} {136}},\
  \bibinfo {pages} {17004} (\bibinfo {year} {2021})}\BibitemShut {NoStop}%
\bibitem [{\citenamefont {Xie}, \citenamefont {Efetov},\ and\ \citenamefont
  {Law}(2022)}]{xie2022valleypolarized}%
  \BibitemOpen
  \bibfield  {author} {\bibinfo {author} {\bibfnamefont {Y.-M.}\ \bibnamefont
  {Xie}}, \bibinfo {author} {\bibfnamefont {D.~K.}\ \bibnamefont {Efetov}}, \
  and\ \bibinfo {author} {\bibfnamefont {K.~T.}\ \bibnamefont {Law}},\ }\href
  {\doibase 10.48550/arXiv.2202.05663} {\enquote {\bibinfo {title}
  {Valley-{Polarized State I}nduced $\varphi_0$-{Josephson Junction in Twisted
  Bilayer G}raphene},}\ } (\bibinfo {year} {2022}),\ \Eprint
  {http://arxiv.org/abs/2202.05663} {arXiv:2202.05663 [cond-mat.mes-hall]}
  \BibitemShut {NoStop}%
\bibitem [{\citenamefont {Pal}\ \emph {et~al.}(2022)\citenamefont {Pal},
  \citenamefont {Chakraborty}, \citenamefont {Sivakumar}, \citenamefont
  {Davydova}, \citenamefont {Gopi}, \citenamefont {Pandeya}, \citenamefont
  {Krieger}, \citenamefont {Zhang}, \citenamefont {Ju}, \citenamefont {Yuan}
  \emph {et~al.}}]{pal2022josephson}%
  \BibitemOpen
  \bibfield  {author} {\bibinfo {author} {\bibfnamefont {B.}~\bibnamefont
  {Pal}}, \bibinfo {author} {\bibfnamefont {A.}~\bibnamefont {Chakraborty}},
  \bibinfo {author} {\bibfnamefont {P.~K.}\ \bibnamefont {Sivakumar}}, \bibinfo
  {author} {\bibfnamefont {M.}~\bibnamefont {Davydova}}, \bibinfo {author}
  {\bibfnamefont {A.~K.}\ \bibnamefont {Gopi}}, \bibinfo {author}
  {\bibfnamefont {A.~K.}\ \bibnamefont {Pandeya}}, \bibinfo {author}
  {\bibfnamefont {J.~A.}\ \bibnamefont {Krieger}}, \bibinfo {author}
  {\bibfnamefont {Y.}~\bibnamefont {Zhang}}, \bibinfo {author} {\bibfnamefont
  {S.}~\bibnamefont {Ju}}, \bibinfo {author} {\bibfnamefont {N.}~\bibnamefont
  {Yuan}},  \emph {et~al.},\ }\bibfield  {title} {\enquote {\bibinfo {title}
  {Josephson diode effect from {C}ooper pair momentum in a topological
  semimetal},}\ }\href {\doibase https://doi.org/10.1038/s41567-022-01699-5}
  {\bibfield  {journal} {\bibinfo  {journal} {Nat. Phys.}\ }\textbf {\bibinfo
  {volume} {18}},\ \bibinfo {pages} {1228--1233} (\bibinfo {year}
  {2022})}\BibitemShut {NoStop}%
\bibitem [{\citenamefont {Baumgartner}\ \emph {et~al.}(2022)\citenamefont
  {Baumgartner}, \citenamefont {Fuchs}, \citenamefont {Costa}, \citenamefont
  {Reinhardt}, \citenamefont {Gronin}, \citenamefont {Gardner}, \citenamefont
  {Lindemann}, \citenamefont {Manfra}, \citenamefont {Faria~Junior},
  \citenamefont {Kochan} \emph {et~al.}}]{baumgartner2022supercurrent}%
  \BibitemOpen
  \bibfield  {author} {\bibinfo {author} {\bibfnamefont {C.}~\bibnamefont
  {Baumgartner}}, \bibinfo {author} {\bibfnamefont {L.}~\bibnamefont {Fuchs}},
  \bibinfo {author} {\bibfnamefont {A.}~\bibnamefont {Costa}}, \bibinfo
  {author} {\bibfnamefont {S.}~\bibnamefont {Reinhardt}}, \bibinfo {author}
  {\bibfnamefont {S.}~\bibnamefont {Gronin}}, \bibinfo {author} {\bibfnamefont
  {G.~C.}\ \bibnamefont {Gardner}}, \bibinfo {author} {\bibfnamefont
  {T.}~\bibnamefont {Lindemann}}, \bibinfo {author} {\bibfnamefont {M.~J.}\
  \bibnamefont {Manfra}}, \bibinfo {author} {\bibfnamefont {P.~E.}\
  \bibnamefont {Faria~Junior}}, \bibinfo {author} {\bibfnamefont
  {D.}~\bibnamefont {Kochan}},  \emph {et~al.},\ }\bibfield  {title} {\enquote
  {\bibinfo {title} {Supercurrent rectification and magnetochiral effects in
  symmetric {J}osephson junctions},}\ }\href {\doibase
  https://doi.org/10.1038/s41565-021-01009-9} {\bibfield  {journal} {\bibinfo
  {journal} {Nat. Nanotechnol.}\ }\textbf {\bibinfo {volume} {17}},\ \bibinfo
  {pages} {39--44} (\bibinfo {year} {2022})}\BibitemShut {NoStop}%
\bibitem [{\citenamefont {Wu}\ \emph {et~al.}(2022)\citenamefont {Wu},
  \citenamefont {Wang}, \citenamefont {Xu}, \citenamefont {Sivakumar},
  \citenamefont {Pasco}, \citenamefont {Filippozzi}, \citenamefont {Parkin},
  \citenamefont {Zeng}, \citenamefont {McQueen},\ and\ \citenamefont
  {Ali}}]{wu2022field}%
  \BibitemOpen
  \bibfield  {author} {\bibinfo {author} {\bibfnamefont {H.}~\bibnamefont
  {Wu}}, \bibinfo {author} {\bibfnamefont {Y.}~\bibnamefont {Wang}}, \bibinfo
  {author} {\bibfnamefont {Y.}~\bibnamefont {Xu}}, \bibinfo {author}
  {\bibfnamefont {P.~K.}\ \bibnamefont {Sivakumar}}, \bibinfo {author}
  {\bibfnamefont {C.}~\bibnamefont {Pasco}}, \bibinfo {author} {\bibfnamefont
  {U.}~\bibnamefont {Filippozzi}}, \bibinfo {author} {\bibfnamefont {S.~S.}\
  \bibnamefont {Parkin}}, \bibinfo {author} {\bibfnamefont {Y.-J.}\
  \bibnamefont {Zeng}}, \bibinfo {author} {\bibfnamefont {T.}~\bibnamefont
  {McQueen}}, \ and\ \bibinfo {author} {\bibfnamefont {M.~N.}\ \bibnamefont
  {Ali}},\ }\bibfield  {title} {\enquote {\bibinfo {title} {The field-free
  {J}osephson diode in a van der {W}aals heterostructure},}\ }\href {\doibase
  https://doi.org/10.1038/s41586-022-04504-8} {\bibfield  {journal} {\bibinfo
  {journal} {Nature}\ }\textbf {\bibinfo {volume} {604}},\ \bibinfo {pages}
  {653--656} (\bibinfo {year} {2022})}\BibitemShut {NoStop}%
\bibitem [{\citenamefont {Golod}\ and\ \citenamefont
  {Krasnov}(2022)}]{Golod2022Demonstration}%
  \BibitemOpen
  \bibfield  {author} {\bibinfo {author} {\bibfnamefont {T.}~\bibnamefont
  {Golod}}\ and\ \bibinfo {author} {\bibfnamefont {V.~M.}\ \bibnamefont
  {Krasnov}},\ }\bibfield  {title} {\enquote {\bibinfo {title} {Demonstration
  of a superconducting diode-with-memory, operational at zero magnetic field
  with switchable nonreciprocity},}\ }\href {\doibase
  10.1038/s41467-022-31256-w} {\bibfield  {journal} {\bibinfo  {journal} {Nat.
  Commun.}\ }\textbf {\bibinfo {volume} {13}},\ \bibinfo {pages} {3658}
  (\bibinfo {year} {2022})}\BibitemShut {NoStop}%
\bibitem [{\citenamefont {Jeon}\ \emph {et~al.}(2022)\citenamefont {Jeon},
  \citenamefont {Kim}, \citenamefont {Yoon}, \citenamefont {Jeon},
  \citenamefont {Han}, \citenamefont {Cottet}, \citenamefont {Kontos},\ and\
  \citenamefont {Parkin}}]{jeon2022zero}%
  \BibitemOpen
  \bibfield  {author} {\bibinfo {author} {\bibfnamefont {K.-R.}\ \bibnamefont
  {Jeon}}, \bibinfo {author} {\bibfnamefont {J.-K.}\ \bibnamefont {Kim}},
  \bibinfo {author} {\bibfnamefont {J.}~\bibnamefont {Yoon}}, \bibinfo {author}
  {\bibfnamefont {J.-C.}\ \bibnamefont {Jeon}}, \bibinfo {author}
  {\bibfnamefont {H.}~\bibnamefont {Han}}, \bibinfo {author} {\bibfnamefont
  {A.}~\bibnamefont {Cottet}}, \bibinfo {author} {\bibfnamefont
  {T.}~\bibnamefont {Kontos}}, \ and\ \bibinfo {author} {\bibfnamefont {S.~S.}\
  \bibnamefont {Parkin}},\ }\bibfield  {title} {\enquote {\bibinfo {title}
  {Zero-field polarity-reversible {J}osephson supercurrent diodes enabled by a
  proximity-magnetized {P}t barrier},}\ }\href {\doibase
  https://doi.org/10.1038/s41563-022-01300-7} {\bibfield  {journal} {\bibinfo
  {journal} {Nat. Mater.}\ }\textbf {\bibinfo {volume} {21}},\ \bibinfo {pages}
  {1008--1013} (\bibinfo {year} {2022})}\BibitemShut {NoStop}%
\bibitem [{\citenamefont {Turini}\ \emph {et~al.}(2022)\citenamefont {Turini},
  \citenamefont {Salimian}, \citenamefont {Carrega}, \citenamefont {Iorio},
  \citenamefont {Strambini}, \citenamefont {Giazotto}, \citenamefont {Zannier},
  \citenamefont {Sorba},\ and\ \citenamefont
  {Heun}}]{doi:10.1021/acs.nanolett.2c02899}%
  \BibitemOpen
  \bibfield  {author} {\bibinfo {author} {\bibfnamefont {B.}~\bibnamefont
  {Turini}}, \bibinfo {author} {\bibfnamefont {S.}~\bibnamefont {Salimian}},
  \bibinfo {author} {\bibfnamefont {M.}~\bibnamefont {Carrega}}, \bibinfo
  {author} {\bibfnamefont {A.}~\bibnamefont {Iorio}}, \bibinfo {author}
  {\bibfnamefont {E.}~\bibnamefont {Strambini}}, \bibinfo {author}
  {\bibfnamefont {F.}~\bibnamefont {Giazotto}}, \bibinfo {author}
  {\bibfnamefont {V.}~\bibnamefont {Zannier}}, \bibinfo {author} {\bibfnamefont
  {L.}~\bibnamefont {Sorba}}, \ and\ \bibinfo {author} {\bibfnamefont
  {S.}~\bibnamefont {Heun}},\ }\bibfield  {title} {\enquote {\bibinfo {title}
  {Josephson {D}iode {E}ffect in {H}igh-{M}obility {InSb N}anoflags},}\ }\href
  {\doibase 10.1021/acs.nanolett.2c02899} {\bibfield  {journal} {\bibinfo
  {journal} {Nano Lett.}\ }\textbf {\bibinfo {volume} {22}},\ \bibinfo {pages}
  {8502--8508} (\bibinfo {year} {2022})}\BibitemShut {NoStop}%
\bibitem [{\citenamefont {Daido}, \citenamefont {Ikeda},\ and\ \citenamefont
  {Yanase}(2022)}]{daido2022intrinsic}%
  \BibitemOpen
  \bibfield  {author} {\bibinfo {author} {\bibfnamefont {A.}~\bibnamefont
  {Daido}}, \bibinfo {author} {\bibfnamefont {Y.}~\bibnamefont {Ikeda}}, \ and\
  \bibinfo {author} {\bibfnamefont {Y.}~\bibnamefont {Yanase}},\ }\bibfield
  {title} {\enquote {\bibinfo {title} {Intrinsic {S}uperconducting {D}iode
  {E}ffect},}\ }\href {\doibase 10.1103/PhysRevLett.128.037001} {\bibfield
  {journal} {\bibinfo  {journal} {Phys. Rev. Lett.}\ }\textbf {\bibinfo
  {volume} {128}},\ \bibinfo {pages} {037001} (\bibinfo {year}
  {2022})}\BibitemShut {NoStop}%
\bibitem [{\citenamefont {Yuan}\ and\ \citenamefont
  {Fu}(2022)}]{yuan2022supercurrent}%
  \BibitemOpen
  \bibfield  {author} {\bibinfo {author} {\bibfnamefont {N.~F.}\ \bibnamefont
  {Yuan}}\ and\ \bibinfo {author} {\bibfnamefont {L.}~\bibnamefont {Fu}},\
  }\bibfield  {title} {\enquote {\bibinfo {title} {Supercurrent diode effect
  and finite-momentum superconductors},}\ }\href {\doibase
  https://doi.org/10.1073/pnas.2119548119} {\bibfield  {journal} {\bibinfo
  {journal} {Proc. Natl. Acad. Sci.}\ }\textbf {\bibinfo {volume} {119}},\
  \bibinfo {pages} {e2119548119} (\bibinfo {year} {2022})}\BibitemShut
  {NoStop}%
\bibitem [{\citenamefont {Karabassov}\ \emph {et~al.}(2022)\citenamefont
  {Karabassov}, \citenamefont {Bobkova}, \citenamefont {Golubov},\ and\
  \citenamefont {Vasenko}}]{PhysRevB.106.224509}%
  \BibitemOpen
  \bibfield  {author} {\bibinfo {author} {\bibfnamefont {T.}~\bibnamefont
  {Karabassov}}, \bibinfo {author} {\bibfnamefont {I.~V.}\ \bibnamefont
  {Bobkova}}, \bibinfo {author} {\bibfnamefont {A.~A.}\ \bibnamefont
  {Golubov}}, \ and\ \bibinfo {author} {\bibfnamefont {A.~S.}\ \bibnamefont
  {Vasenko}},\ }\bibfield  {title} {\enquote {\bibinfo {title} {Hybrid helical
  state and superconducting diode effect in
  superconductor/ferromagnet/topological insulator heterostructures},}\ }\href
  {\doibase 10.1103/PhysRevB.106.224509} {\bibfield  {journal} {\bibinfo
  {journal} {Phys. Rev. B}\ }\textbf {\bibinfo {volume} {106}},\ \bibinfo
  {pages} {224509} (\bibinfo {year} {2022})}\BibitemShut {NoStop}%
\bibitem [{\citenamefont {Zhai}\ \emph {et~al.}(2022)\citenamefont {Zhai},
  \citenamefont {Li}, \citenamefont {Wen}, \citenamefont {Wu},\ and\
  \citenamefont {He}}]{PhysRevB.106.L140505}%
  \BibitemOpen
  \bibfield  {author} {\bibinfo {author} {\bibfnamefont {B.}~\bibnamefont
  {Zhai}}, \bibinfo {author} {\bibfnamefont {B.}~\bibnamefont {Li}}, \bibinfo
  {author} {\bibfnamefont {Y.}~\bibnamefont {Wen}}, \bibinfo {author}
  {\bibfnamefont {F.}~\bibnamefont {Wu}}, \ and\ \bibinfo {author}
  {\bibfnamefont {J.}~\bibnamefont {He}},\ }\bibfield  {title} {\enquote
  {\bibinfo {title} {Prediction of ferroelectric superconductors with
  reversible superconducting diode effect},}\ }\href {\doibase
  10.1103/PhysRevB.106.L140505} {\bibfield  {journal} {\bibinfo  {journal}
  {Phys. Rev. B}\ }\textbf {\bibinfo {volume} {106}},\ \bibinfo {pages}
  {L140505} (\bibinfo {year} {2022})}\BibitemShut {NoStop}%
\bibitem [{\citenamefont {Daido}\ and\ \citenamefont
  {Yanase}(2022)}]{PhysRevB.106.205206}%
  \BibitemOpen
  \bibfield  {author} {\bibinfo {author} {\bibfnamefont {A.}~\bibnamefont
  {Daido}}\ and\ \bibinfo {author} {\bibfnamefont {Y.}~\bibnamefont {Yanase}},\
  }\bibfield  {title} {\enquote {\bibinfo {title} {Superconducting diode effect
  and nonreciprocal transition lines},}\ }\href {\doibase
  10.1103/PhysRevB.106.205206} {\bibfield  {journal} {\bibinfo  {journal}
  {Phys. Rev. B}\ }\textbf {\bibinfo {volume} {106}},\ \bibinfo {pages}
  {205206} (\bibinfo {year} {2022})}\BibitemShut {NoStop}%
\bibitem [{\citenamefont {Scammell}, \citenamefont {Li},\ and\ \citenamefont
  {Scheurer}(2022)}]{Scammell_2022}%
  \BibitemOpen
  \bibfield  {author} {\bibinfo {author} {\bibfnamefont {H.~D.}\ \bibnamefont
  {Scammell}}, \bibinfo {author} {\bibfnamefont {J.~I.~A.}\ \bibnamefont {Li}},
  \ and\ \bibinfo {author} {\bibfnamefont {M.~S.}\ \bibnamefont {Scheurer}},\
  }\bibfield  {title} {\enquote {\bibinfo {title} {Theory of zero-field
  superconducting diode effect in twisted trilayer graphene},}\ }\href
  {\doibase 10.1088/2053-1583/ac5b16} {\bibfield  {journal} {\bibinfo
  {journal} {2D Materials}\ }\textbf {\bibinfo {volume} {9}},\ \bibinfo {pages}
  {025027} (\bibinfo {year} {2022})}\BibitemShut {NoStop}%
\bibitem [{\citenamefont {Wang}, \citenamefont {Wang},\ and\ \citenamefont
  {Wu}(2022)}]{wang2022symmetry}%
  \BibitemOpen
  \bibfield  {author} {\bibinfo {author} {\bibfnamefont {D.}~\bibnamefont
  {Wang}}, \bibinfo {author} {\bibfnamefont {Q.-H.}\ \bibnamefont {Wang}}, \
  and\ \bibinfo {author} {\bibfnamefont {C.}~\bibnamefont {Wu}},\ }\href
  {\doibase 10.48550/arXiv.2209.12646} {\enquote {\bibinfo {title} {Symmetry
  constraints on direct-current josephson diodes},}\ } (\bibinfo {year}
  {2022}),\ \Eprint {http://arxiv.org/abs/2209.12646} {arXiv:2209.12646
  [cond-mat.supr-con]} \BibitemShut {NoStop}%
\bibitem [{\citenamefont {Ando}\ \emph {et~al.}(2020)\citenamefont {Ando},
  \citenamefont {Miyasaka}, \citenamefont {Li}, \citenamefont {Ishizuka},
  \citenamefont {Arakawa}, \citenamefont {Shiota}, \citenamefont {Moriyama},
  \citenamefont {Yanase},\ and\ \citenamefont {Ono}}]{ando2020observation}%
  \BibitemOpen
  \bibfield  {author} {\bibinfo {author} {\bibfnamefont {F.}~\bibnamefont
  {Ando}}, \bibinfo {author} {\bibfnamefont {Y.}~\bibnamefont {Miyasaka}},
  \bibinfo {author} {\bibfnamefont {T.}~\bibnamefont {Li}}, \bibinfo {author}
  {\bibfnamefont {J.}~\bibnamefont {Ishizuka}}, \bibinfo {author}
  {\bibfnamefont {T.}~\bibnamefont {Arakawa}}, \bibinfo {author} {\bibfnamefont
  {Y.}~\bibnamefont {Shiota}}, \bibinfo {author} {\bibfnamefont
  {T.}~\bibnamefont {Moriyama}}, \bibinfo {author} {\bibfnamefont
  {Y.}~\bibnamefont {Yanase}}, \ and\ \bibinfo {author} {\bibfnamefont
  {T.}~\bibnamefont {Ono}},\ }\bibfield  {title} {\enquote {\bibinfo {title}
  {Observation of superconducting diode effect},}\ }\href@noop {} {\bibfield
  {journal} {\bibinfo  {journal} {Nature}\ }\textbf {\bibinfo {volume} {584}},\
  \bibinfo {pages} {373--376} (\bibinfo {year} {2020})}\BibitemShut {NoStop}%
\bibitem [{\citenamefont {Wakatsuki}\ \emph {et~al.}(2017)\citenamefont
  {Wakatsuki}, \citenamefont {Saito}, \citenamefont {Hoshino}, \citenamefont
  {Itahashi}, \citenamefont {Ideue}, \citenamefont {Ezawa}, \citenamefont
  {Iwasa},\ and\ \citenamefont {Nagaosa}}]{wakatsuki2017nonreciprocal}%
  \BibitemOpen
  \bibfield  {author} {\bibinfo {author} {\bibfnamefont {R.}~\bibnamefont
  {Wakatsuki}}, \bibinfo {author} {\bibfnamefont {Y.}~\bibnamefont {Saito}},
  \bibinfo {author} {\bibfnamefont {S.}~\bibnamefont {Hoshino}}, \bibinfo
  {author} {\bibfnamefont {Y.~M.}\ \bibnamefont {Itahashi}}, \bibinfo {author}
  {\bibfnamefont {T.}~\bibnamefont {Ideue}}, \bibinfo {author} {\bibfnamefont
  {M.}~\bibnamefont {Ezawa}}, \bibinfo {author} {\bibfnamefont
  {Y.}~\bibnamefont {Iwasa}}, \ and\ \bibinfo {author} {\bibfnamefont
  {N.}~\bibnamefont {Nagaosa}},\ }\bibfield  {title} {\enquote {\bibinfo
  {title} {Nonreciprocal charge transport in noncentrosymmetric
  superconductors},}\ }\href {\doibase https://doi.org/10.1126/sciadv.1602390}
  {\bibfield  {journal} {\bibinfo  {journal} {Sci. Adv.}\ }\textbf {\bibinfo
  {volume} {3}},\ \bibinfo {pages} {e1602390} (\bibinfo {year}
  {2017})}\BibitemShut {NoStop}%
\bibitem [{\citenamefont {Zhang}\ \emph {et~al.}(2020)\citenamefont {Zhang},
  \citenamefont {Xu}, \citenamefont {Zou}, \citenamefont {Ai}, \citenamefont
  {Dong}, \citenamefont {Huang}, \citenamefont {Leng}, \citenamefont {Liu},
  \citenamefont {Zhang}, \citenamefont {Jia} \emph
  {et~al.}}]{zhang2020nonreciprocal}%
  \BibitemOpen
  \bibfield  {author} {\bibinfo {author} {\bibfnamefont {E.}~\bibnamefont
  {Zhang}}, \bibinfo {author} {\bibfnamefont {X.}~\bibnamefont {Xu}}, \bibinfo
  {author} {\bibfnamefont {Y.-C.}\ \bibnamefont {Zou}}, \bibinfo {author}
  {\bibfnamefont {L.}~\bibnamefont {Ai}}, \bibinfo {author} {\bibfnamefont
  {X.}~\bibnamefont {Dong}}, \bibinfo {author} {\bibfnamefont {C.}~\bibnamefont
  {Huang}}, \bibinfo {author} {\bibfnamefont {P.}~\bibnamefont {Leng}},
  \bibinfo {author} {\bibfnamefont {S.}~\bibnamefont {Liu}}, \bibinfo {author}
  {\bibfnamefont {Y.}~\bibnamefont {Zhang}}, \bibinfo {author} {\bibfnamefont
  {Z.}~\bibnamefont {Jia}},  \emph {et~al.},\ }\bibfield  {title} {\enquote
  {\bibinfo {title} {Nonreciprocal superconducting {NbS}e2 antenna},}\ }\href
  {\doibase https://doi.org/10.1038/s41467-020-19459-5} {\bibfield  {journal}
  {\bibinfo  {journal} {Nat. Commun.}\ }\textbf {\bibinfo {volume} {11}},\
  \bibinfo {pages} {1--9} (\bibinfo {year} {2020})}\BibitemShut {NoStop}%
\bibitem [{\citenamefont {Narita}\ \emph {et~al.}(2022)\citenamefont {Narita},
  \citenamefont {Ishizuka}, \citenamefont {Kawarazaki}, \citenamefont {Kan},
  \citenamefont {Shiota}, \citenamefont {Moriyama}, \citenamefont {Shimakawa},
  \citenamefont {Ognev}, \citenamefont {Samardak}, \citenamefont {Yanase},\
  and\ \citenamefont {Ono}}]{Narita2022Field}%
  \BibitemOpen
  \bibfield  {author} {\bibinfo {author} {\bibfnamefont {H.}~\bibnamefont
  {Narita}}, \bibinfo {author} {\bibfnamefont {J.}~\bibnamefont {Ishizuka}},
  \bibinfo {author} {\bibfnamefont {R.}~\bibnamefont {Kawarazaki}}, \bibinfo
  {author} {\bibfnamefont {D.}~\bibnamefont {Kan}}, \bibinfo {author}
  {\bibfnamefont {Y.}~\bibnamefont {Shiota}}, \bibinfo {author} {\bibfnamefont
  {T.}~\bibnamefont {Moriyama}}, \bibinfo {author} {\bibfnamefont
  {Y.}~\bibnamefont {Shimakawa}}, \bibinfo {author} {\bibfnamefont {A.~V.}\
  \bibnamefont {Ognev}}, \bibinfo {author} {\bibfnamefont {A.~S.}\ \bibnamefont
  {Samardak}}, \bibinfo {author} {\bibfnamefont {Y.}~\bibnamefont {Yanase}}, \
  and\ \bibinfo {author} {\bibfnamefont {T.}~\bibnamefont {Ono}},\ }\bibfield
  {title} {\enquote {\bibinfo {title} {Field-free superconducting diode effect
  in noncentrosymmetric superconductor/ferromagnet multilayers},}\ }\href
  {\doibase 10.1038/s41565-022-01159-4} {\bibfield  {journal} {\bibinfo
  {journal} {Nat. Nanotechnol.}\ }\textbf {\bibinfo {volume} {17}},\ \bibinfo
  {pages} {823—828} (\bibinfo {year} {2022})}\BibitemShut {NoStop}%
\bibitem [{\citenamefont {Lin}\ \emph {et~al.}(2022)\citenamefont {Lin},
  \citenamefont {Siriviboon}, \citenamefont {Scammell}, \citenamefont {Liu},
  \citenamefont {Rhodes}, \citenamefont {Watanabe}, \citenamefont {Taniguchi},
  \citenamefont {Hone}, \citenamefont {Scheurer},\ and\ \citenamefont
  {Li}}]{lin2022zero}%
  \BibitemOpen
  \bibfield  {author} {\bibinfo {author} {\bibfnamefont {J.-X.}\ \bibnamefont
  {Lin}}, \bibinfo {author} {\bibfnamefont {P.}~\bibnamefont {Siriviboon}},
  \bibinfo {author} {\bibfnamefont {H.~D.}\ \bibnamefont {Scammell}}, \bibinfo
  {author} {\bibfnamefont {S.}~\bibnamefont {Liu}}, \bibinfo {author}
  {\bibfnamefont {D.}~\bibnamefont {Rhodes}}, \bibinfo {author} {\bibfnamefont
  {K.}~\bibnamefont {Watanabe}}, \bibinfo {author} {\bibfnamefont
  {T.}~\bibnamefont {Taniguchi}}, \bibinfo {author} {\bibfnamefont
  {J.}~\bibnamefont {Hone}}, \bibinfo {author} {\bibfnamefont {M.~S.}\
  \bibnamefont {Scheurer}}, \ and\ \bibinfo {author} {\bibfnamefont
  {J.}~\bibnamefont {Li}},\ }\bibfield  {title} {\enquote {\bibinfo {title}
  {Zero-field superconducting diode effect in small-twist-angle trilayer
  graphene},}\ }\href {\doibase https://doi.org/10.1038/s41567-022-01700-1}
  {\bibfield  {journal} {\bibinfo  {journal} {Nature Physics}\ }\textbf
  {\bibinfo {volume} {18}},\ \bibinfo {pages} {1221--1227} (\bibinfo {year}
  {2022})}\BibitemShut {NoStop}%
\bibitem [{\citenamefont {Hou}\ \emph {et~al.}(2022)\citenamefont {Hou},
  \citenamefont {Nichele}, \citenamefont {Chi}, \citenamefont {Lodesani},
  \citenamefont {Wu}, \citenamefont {Ritter}, \citenamefont {Haxell},
  \citenamefont {Davydova}, \citenamefont {Ili{\'c}}, \citenamefont {Bergeret}
  \emph {et~al.}}]{hou2022ubiquitous}%
  \BibitemOpen
  \bibfield  {author} {\bibinfo {author} {\bibfnamefont {Y.}~\bibnamefont
  {Hou}}, \bibinfo {author} {\bibfnamefont {F.}~\bibnamefont {Nichele}},
  \bibinfo {author} {\bibfnamefont {H.}~\bibnamefont {Chi}}, \bibinfo {author}
  {\bibfnamefont {A.}~\bibnamefont {Lodesani}}, \bibinfo {author}
  {\bibfnamefont {Y.}~\bibnamefont {Wu}}, \bibinfo {author} {\bibfnamefont
  {M.~F.}\ \bibnamefont {Ritter}}, \bibinfo {author} {\bibfnamefont {D.~Z.}\
  \bibnamefont {Haxell}}, \bibinfo {author} {\bibfnamefont {M.}~\bibnamefont
  {Davydova}}, \bibinfo {author} {\bibfnamefont {S.}~\bibnamefont {Ili{\'c}}},
  \bibinfo {author} {\bibfnamefont {F.~S.}\ \bibnamefont {Bergeret}},  \emph
  {et~al.},\ }\bibfield  {title} {\enquote {\bibinfo {title} {Ubiquitous
  {S}uperconducting {D}iode {E}ffect in {S}uperconductor {T}hin {F}ilms},}\
  }\href {\doibase https://doi.org/10.48550/arXiv.2205.09276} {\bibfield
  {journal} {\bibinfo  {journal} {arXiv:2205.09276}\ } (\bibinfo {year}
  {2022}),\ https://doi.org/10.48550/arXiv.2205.09276}\BibitemShut {NoStop}%
\bibitem [{\citenamefont {Suri}\ \emph {et~al.}(2022)\citenamefont {Suri},
  \citenamefont {Kamra}, \citenamefont {Meier}, \citenamefont {Kronseder},
  \citenamefont {Belzig}, \citenamefont {Back},\ and\ \citenamefont
  {Strunk}}]{Suri2022Non}%
  \BibitemOpen
  \bibfield  {author} {\bibinfo {author} {\bibfnamefont {D.}~\bibnamefont
  {Suri}}, \bibinfo {author} {\bibfnamefont {A.}~\bibnamefont {Kamra}},
  \bibinfo {author} {\bibfnamefont {T.~N.~G.}\ \bibnamefont {Meier}}, \bibinfo
  {author} {\bibfnamefont {M.}~\bibnamefont {Kronseder}}, \bibinfo {author}
  {\bibfnamefont {W.}~\bibnamefont {Belzig}}, \bibinfo {author} {\bibfnamefont
  {C.~H.}\ \bibnamefont {Back}}, \ and\ \bibinfo {author} {\bibfnamefont
  {C.}~\bibnamefont {Strunk}},\ }\bibfield  {title} {\enquote {\bibinfo {title}
  {Non-reciprocity of vortex-limited critical current in conventional
  superconducting micro-bridges},}\ }\href {\doibase 10.1063/5.0109753}
  {\bibfield  {journal} {\bibinfo  {journal} {Appl. Phys. Lett.}\ }\textbf
  {\bibinfo {volume} {121}},\ \bibinfo {pages} {102601} (\bibinfo {year}
  {2022})}\BibitemShut {NoStop}%
\bibitem [{\citenamefont {Ustavschikov}\ \emph {et~al.}(2022)\citenamefont
  {Ustavschikov}, \citenamefont {Levichev}, \citenamefont {Pashenkin},
  \citenamefont {Gusev}, \citenamefont {Gusev},\ and\ \citenamefont
  {Vodolazov}}]{ustavschikov2022diode}%
  \BibitemOpen
  \bibfield  {author} {\bibinfo {author} {\bibfnamefont {S.}~\bibnamefont
  {Ustavschikov}}, \bibinfo {author} {\bibfnamefont {M.~Y.}\ \bibnamefont
  {Levichev}}, \bibinfo {author} {\bibfnamefont {I.~Y.}\ \bibnamefont
  {Pashenkin}}, \bibinfo {author} {\bibfnamefont {N.}~\bibnamefont {Gusev}},
  \bibinfo {author} {\bibfnamefont {S.}~\bibnamefont {Gusev}}, \ and\ \bibinfo
  {author} {\bibfnamefont {D.~Y.}\ \bibnamefont {Vodolazov}},\ }\bibfield
  {title} {\enquote {\bibinfo {title} {Diode {Effect in a Superconducting
  Hybrid Cu/MoN Strip with a Lateral C}ut},}\ }\href {\doibase
  https://doi.org/10.1134/S1063776122080064} {\bibfield  {journal} {\bibinfo
  {journal} {J. Exp. Theor. Phys.}\ }\textbf {\bibinfo {volume} {135}},\
  \bibinfo {pages} {226--230} (\bibinfo {year} {2022})}\BibitemShut {NoStop}%
\bibitem [{\citenamefont {Gutfreund}\ \emph {et~al.}(2023)\citenamefont
  {Gutfreund}, \citenamefont {Matsuki}, \citenamefont {Plastovets},
  \citenamefont {Noah}, \citenamefont {Gorzawski}, \citenamefont {Fridman},
  \citenamefont {Yang}, \citenamefont {Buzdin}, \citenamefont {Millo},
  \citenamefont {Robinson} \emph {et~al.}}]{gutfreund2023direct}%
  \BibitemOpen
  \bibfield  {author} {\bibinfo {author} {\bibfnamefont {A.}~\bibnamefont
  {Gutfreund}}, \bibinfo {author} {\bibfnamefont {H.}~\bibnamefont {Matsuki}},
  \bibinfo {author} {\bibfnamefont {V.}~\bibnamefont {Plastovets}}, \bibinfo
  {author} {\bibfnamefont {A.}~\bibnamefont {Noah}}, \bibinfo {author}
  {\bibfnamefont {L.}~\bibnamefont {Gorzawski}}, \bibinfo {author}
  {\bibfnamefont {N.}~\bibnamefont {Fridman}}, \bibinfo {author} {\bibfnamefont
  {G.}~\bibnamefont {Yang}}, \bibinfo {author} {\bibfnamefont {A.}~\bibnamefont
  {Buzdin}}, \bibinfo {author} {\bibfnamefont {O.}~\bibnamefont {Millo}},
  \bibinfo {author} {\bibfnamefont {J.~W.}\ \bibnamefont {Robinson}},  \emph
  {et~al.},\ }\bibfield  {title} {\enquote {\bibinfo {title} {Direct
  {O}bservation of a {S}uperconducting {V}ortex {D}iode},}\ }\href {\doibase
  10.1038/s41467-023-37294-2} {\bibfield  {journal} {\bibinfo  {journal} {Nat.
  Commun.}\ }\textbf {\bibinfo {volume} {14}},\ \bibinfo {pages} {1630}
  (\bibinfo {year} {2023})}\BibitemShut {NoStop}%
\bibitem [{\citenamefont {Vodolazov}\ and\ \citenamefont
  {Peeters}(2005)}]{PhysRevB.72.172508}%
  \BibitemOpen
  \bibfield  {author} {\bibinfo {author} {\bibfnamefont {D.~Y.}\ \bibnamefont
  {Vodolazov}}\ and\ \bibinfo {author} {\bibfnamefont {F.~M.}\ \bibnamefont
  {Peeters}},\ }\bibfield  {title} {\enquote {\bibinfo {title} {Superconducting
  rectifier based on the asymmetric surface barrier effect},}\ }\href {\doibase
  10.1103/PhysRevB.72.172508} {\bibfield  {journal} {\bibinfo  {journal} {Phys.
  Rev. B}\ }\textbf {\bibinfo {volume} {72}},\ \bibinfo {pages} {172508}
  (\bibinfo {year} {2005})}\BibitemShut {NoStop}%
\bibitem [{Roy()}]{Royce}%
  \BibitemOpen
  \href@noop {} {}\bibinfo {note} {The Royce Deposition System is a
  multi-chamber, multi-technique thin film deposition tool based at the
  University of Leeds as part of the \href{https://www.royce.ac.uk/}{Henry
  Royce Institute}.}\BibitemShut {Stop}%
\bibitem [{\citenamefont {Quarterman}\ \emph {et~al.}(2020)\citenamefont
  {Quarterman}, \citenamefont {Satchell}, \citenamefont {Kirby}, \citenamefont
  {Loloee}, \citenamefont {Burnell}, \citenamefont {Birge},\ and\ \citenamefont
  {Borchers}}]{PhysRevMaterials.4.074801}%
  \BibitemOpen
  \bibfield  {author} {\bibinfo {author} {\bibfnamefont {P.}~\bibnamefont
  {Quarterman}}, \bibinfo {author} {\bibfnamefont {N.}~\bibnamefont
  {Satchell}}, \bibinfo {author} {\bibfnamefont {B.~J.}\ \bibnamefont {Kirby}},
  \bibinfo {author} {\bibfnamefont {R.}~\bibnamefont {Loloee}}, \bibinfo
  {author} {\bibfnamefont {G.}~\bibnamefont {Burnell}}, \bibinfo {author}
  {\bibfnamefont {N.~O.}\ \bibnamefont {Birge}}, \ and\ \bibinfo {author}
  {\bibfnamefont {J.~A.}\ \bibnamefont {Borchers}},\ }\bibfield  {title}
  {\enquote {\bibinfo {title} {Distortions to the penetration depth and
  coherence length of superconductor/normal-metal superlattices},}\ }\href
  {\doibase 10.1103/PhysRevMaterials.4.074801} {\bibfield  {journal} {\bibinfo
  {journal} {Phys. Rev. Materials}\ }\textbf {\bibinfo {volume} {4}},\ \bibinfo
  {pages} {074801} (\bibinfo {year} {2020})}\BibitemShut {NoStop}%
\bibitem [{\citenamefont {Suter}\ \emph {et~al.}(2005)\citenamefont {Suter},
  \citenamefont {Morenzoni}, \citenamefont {Garifianov}, \citenamefont
  {Khasanov}, \citenamefont {Kirk}, \citenamefont {Luetkens}, \citenamefont
  {Prokscha},\ and\ \citenamefont {Horisberger}}]{PhysRevB.72.024506}%
  \BibitemOpen
  \bibfield  {author} {\bibinfo {author} {\bibfnamefont {A.}~\bibnamefont
  {Suter}}, \bibinfo {author} {\bibfnamefont {E.}~\bibnamefont {Morenzoni}},
  \bibinfo {author} {\bibfnamefont {N.}~\bibnamefont {Garifianov}}, \bibinfo
  {author} {\bibfnamefont {R.}~\bibnamefont {Khasanov}}, \bibinfo {author}
  {\bibfnamefont {E.}~\bibnamefont {Kirk}}, \bibinfo {author} {\bibfnamefont
  {H.}~\bibnamefont {Luetkens}}, \bibinfo {author} {\bibfnamefont
  {T.}~\bibnamefont {Prokscha}}, \ and\ \bibinfo {author} {\bibfnamefont
  {M.}~\bibnamefont {Horisberger}},\ }\bibfield  {title} {\enquote {\bibinfo
  {title} {Observation of nonexponential magnetic penetration profiles in the
  {M}eissner state: {A} manifestation of nonlocal effects in
  superconductors},}\ }\href {\doibase 10.1103/PhysRevB.72.024506} {\bibfield
  {journal} {\bibinfo  {journal} {Phys. Rev. B}\ }\textbf {\bibinfo {volume}
  {72}},\ \bibinfo {pages} {024506} (\bibinfo {year} {2005})}\BibitemShut
  {NoStop}%
\bibitem [{\citenamefont {Il’in}\ \emph {et~al.}(2010)\citenamefont
  {Il’in}, \citenamefont {Rall}, \citenamefont {Siegel}, \citenamefont
  {Engel}, \citenamefont {Schilling}, \citenamefont {Semenov},\ and\
  \citenamefont {Huebers}}]{ILIN2010953}%
  \BibitemOpen
  \bibfield  {author} {\bibinfo {author} {\bibfnamefont {K.}~\bibnamefont
  {Il’in}}, \bibinfo {author} {\bibfnamefont {D.}~\bibnamefont {Rall}},
  \bibinfo {author} {\bibfnamefont {M.}~\bibnamefont {Siegel}}, \bibinfo
  {author} {\bibfnamefont {A.}~\bibnamefont {Engel}}, \bibinfo {author}
  {\bibfnamefont {A.}~\bibnamefont {Schilling}}, \bibinfo {author}
  {\bibfnamefont {A.}~\bibnamefont {Semenov}}, \ and\ \bibinfo {author}
  {\bibfnamefont {H.-W.}\ \bibnamefont {Huebers}},\ }\bibfield  {title}
  {\enquote {\bibinfo {title} {Influence of thickness, width and temperature on
  critical current density of nb thin film structures},}\ }\href {\doibase
  https://doi.org/10.1016/j.physc.2010.02.042} {\bibfield  {journal} {\bibinfo
  {journal} {Phys. C: Supercond. Appl.}\ }\textbf {\bibinfo {volume} {470}},\
  \bibinfo {pages} {953--956} (\bibinfo {year} {2010})},\ \bibinfo {note}
  {vortex Matter in Nanostructured Superconductors}\BibitemShut {NoStop}%
\bibitem [{\citenamefont {Zhu}, \citenamefont {Lockhart},\ and\ \citenamefont
  {Turneare}(1994)}]{ZHU19941357}%
  \BibitemOpen
  \bibfield  {author} {\bibinfo {author} {\bibfnamefont {J.}~\bibnamefont
  {Zhu}}, \bibinfo {author} {\bibfnamefont {J.}~\bibnamefont {Lockhart}}, \
  and\ \bibinfo {author} {\bibfnamefont {J.}~\bibnamefont {Turneare}},\
  }\bibfield  {title} {\enquote {\bibinfo {title} {Pinning forces in a
  disk-shaped superconducting niobium film},}\ }\href {\doibase
  https://doi.org/10.1016/0921-4526(94)91008-1} {\bibfield  {journal} {\bibinfo
   {journal} {Phys. B: Condens. Matter}\ }\textbf {\bibinfo {volume}
  {194-196}},\ \bibinfo {pages} {1357--1358} (\bibinfo {year}
  {1994})}\BibitemShut {NoStop}%
\end{thebibliography}%

\clearpage

\clearpage

\clearpage

\clearpage

\clearpage

\end{document}